\documentclass[times,preprint,nopreprintline]{elsarticle}
\usepackage{framed,multirow}

\usepackage{amssymb}
\usepackage{amsmath}
\usepackage{latexsym}
\usepackage{booktabs}
\usepackage{graphicx}
\usepackage{subcaption}
\usepackage[pagewise]{lineno}
\graphicspath{{./images/}}
\usepackage{url}
\newcommand{\hl}[1]{#1}

\journal{Journal of Computational Physics}

\begin{document}

\begin{frontmatter}

\title{SPRAY: A smoothed particle radiation hydrodynamics code for \hl{modeling} high intensity laser-plasma interaction\hl{s}}

\author[1]{Min Ki {Jung}}
\author[2]{Hakhyeon {Kim}}
\author[1]{Su-San {Park}}
\author[1]{Eung Soo {Kim}}
\author[1]{Yong-Su {Na}}
\cortext[cor1]{Corresponding author:}
\author[2]{Sang June {Hahn}\corref{cor1}}
\ead{sjhahn@cau.ac.kr}

\address[1]{Department of Nuclear Engineering, Seoul National University, Seoul, South Korea}
\address[2]{Department of Physics, Chung-Ang University, Seoul, South Korea}

\begin{abstract}
%%%
Here we report the development of SPRAY, a massively parallel GPU accelerated, smoothed particle hydrodynamics (SPH)-based, radiation hydrodynamics (RHD) code designed specifically for simulating high intensity laser-plasma interactions. When a target is irradiated by an intense laser, highly complex fluid deformation occurs due to instabilities, which \hl{is} challenging to study numerically. SPRAY is particle-based, mesh-free, and Lagrangian, which addresses numerical issues that posed difficulties to existing methods. Its SPH formulations for RHD governing equations are tailored toward accurate and reliable simulations of laser-target irradiation phenomena\hl{, and are solved via a time-dependent, flux-limited diffusion method.} A new laser energy coupling module, which is based on the Wentzel-Kramers-Brillouin (WKB) approximation, is implemented with a totally mesh-free ray-tracing scheme that is applicable for arbitrary geometry and dimensions. The accuracy and reliability of the code are demonstrated with a series of benchmark problems. To the authors’ knowledge, this is the first attempt to employ SPH method for simulations of laser-plasma interactions in high energy density physics research. Possible expansions to the code, such as laser beam-beam interaction \hl{modeling} and more sophisticated multi-group radiation transport are left for future development.
%%%%
\end{abstract}

\end{frontmatter}

%\linenumbers

%% main text

\section{Introduction}\label{sec:1}

The interaction of high-intensity lasers with plasmas, both experimentally and numerically, is a popular research topic since the early days of high energy density physics (HEDP) and inertial confinement fusion (ICF) research. The simulation of high energy density plasmas, characterized by high pressure (typically greater than $1\ Mbar$), is key to our understanding of stellar interiors and astrophysical systems \cite{Drake2006,Larsen2017,Colvin2013}. Moreover, ICF experiments, which aim to produce nuclear fusion reactions by garnering the potentials of HEDPs, continue to show promising results on the path to realizing fusion energy production \cite{Zylstra2022}.

In laboratory settings, high energy density plasmas, including laser-driven ICF plasmas, are produced by irradiating solid targets with high-intensity lasers, typically with peak intensities exceeding $10^{14}\ W/cm^2$. When such lasers interact with targets, corona plasmas are formed as a result of ablation, and high compression regions are created by shock waves. In addition, a host of instabilities is induced due to steep pressure and density gradients. Ablative Rayleigh-Taylor instabilities (ARTI), for instance, occur in both the ablation front and the boundary between the fuel and the hot-spot, and it is essential to control and mitigate such phenomena as they seriously degrade the ICF performance \cite{Wang2013,Bates2016}.  However, the multi-scale nature of these physical phenomena poses challenges to experimental diagnosis and analysis. Therefore, it is of great interest to study the interactions between high-intensity lasers and plasmas with numerical analysis.

There are a variety of numerical approaches available to examine high-intensity laser-plasma interactions. Eulerian-based methods \cite{Fryxell2000,vanderHolst2011,Gittings2008}, although well-established and numerically matured, pose difficulties in tracking the moving multi-fluid interfaces due to their grid-based nature. In addition, when the coronal plasma expands as ablation progresses, it becomes challenging to keep track of the boundary surface of the plasma. Moreover, adaptive mesh refinement (AMR) schemes, which are necessary to resolve the fine spatial structures of instabilities, are computationally intensive. On the other hand, there exist a number of codes \cite{Ramis2012,Ramis2016,Larsen1994,Marinak2001} that take the Lagrangian approach to avoid these issues. In these approaches, the plasma boundary is trivially tracked as the Lagrangian fluid elements move in space. However, these codes are generally cell-based, meaning that the fluid elements are represented by staggered cells. This severely limits the ability to handle large fluid deformations that are commonplace in hydrodynamic instabilities in the context of HEDPs.

In this paper, an application of particle-based, mesh-free, Lagrangian approach to solve radiation hydrodynamics (RHD) is presented. SPRAY (\textbf{S}moothed \textbf{P}article \textbf{RA}diation h\textbf{Y}drodynamics), a SPH-based RHD code, is developed specifically for high-intensity laser-plasma interaction simulations. It is, to our knowledge, the first SPH code to model such phenomena with detailed physical \hl{modeling} and material data. SPH, first proposed independently by Lucy \cite{Lucy1977} and Gingold and Monaghan \cite{Gingold1977}, is a particle-based hydrodynamics method, as the name suggests. It does not rely on any mesh, grid or cell to numerically solve hydrodynamics governing equations, and it is Lagrangian in nature, as the fluid elements are tracked by particles. The strengths and advantages of SPH come from its ability to handle multi-fluid dynamics with complex deformations. Because of its versatility and robustness, it is being utilized not only by scientific communities but also in industrial applications, as evidenced by the continued publications of a series of review papers \cite{Monaghan1992,Monaghan2005,Liu2008,Springel2010,Liu2010,Monaghan2012,Price2012,Shadloo2016,Lind2020,Sigalotti2021}.

Since SPH was first developed for and applied to astrophysical problems, there \hl{has} been a continued interest in incorporating radiation transport modeling into the SPH formulation. In particular, modeling ionizing radiation has been attempted by numerous studies, mostly by ray-tracing ionizing photons emitted by radiation sources \hl{\mbox{\cite{Kessel-Deynet2000,Susa2006,Hasegawa2010,Altay2008}}}. There also have been studies where the domain is discretized into virtual cones in order to preserve the directed transport \mbox{\cite{Pawlik2008}}. However, these approaches are highly expensive in terms of computation, and require generation and tracking of virtual photon particles. Because of these shortcomings, there also has been another branch in radiation transport modeling in SPH using the flux-limited diffusion approximation derived by \mbox{\cite{Levermore1981}}. The diffusion approximation takes the zeroth moment of the radiation transport equation and is valid when the opacity is not too small. Studies that employ this approximation include \mbox{\cite{Fryer2006,Whitehouse2005,Mayer2007}}. In addition, moment-based approaches with variable Eddington tensor approximation for optically thin regime have been proposed as well \hl{\mbox{\cite{Petkova2010}}}. Recently, there has been a novel study that proposes solving both the zeroth and the first moment simultaneously, dubbed the "two-moment" method \mbox{\cite{Chan2021}}.

Because SPRAY is a RHD code targeting laser-plasma interactions, the physical characteristics of laser-driven plasmas were taken into consideration when designing the radiation transport scheme. Because the optical depth of laser-produced plasma is generally thick, the use of flux-limited diffusion approximation is justified, so we opted to take this approach for the sake of computational efficiency. In addition, because the hydrodynamics of the laser-plasma interactions needed to be resolved at the same time, RHD is solved via operator splitting method where radiation transport is solved implicitly and hydrodynamics is solved explicitly \mbox{\cite{Whitehouse2005,Whitehouse2004,Bassett2021A, Bassett2021B}}. Furthermore, the radiation transport equations are coupled to the three-temperature description of plasma (ion, electron, and radiation) instead of the one- or at most two-temperature description used in previous studies, because of the large discrepancy between ion and electron temperatures during high intensity laser-plasma interactions.

In SPRAY, laser energy deposition in plasma is described by the inverse bremsstrahlung absorption model \mbox{\cite{Johnston1973}}, and it is numerically implemented in the Wentzel-Kramers-Brillouin (WKB) approximation form \mbox{\cite{Eliezer2003}}. \hl{The} WKB approximation is valid when the laser wavelength is shorter than the plasma gradient scale length, and this condition is satisfied for long pulse lasers with pulse duration in the order of nanoseconds, which are the main target of this research \mbox{\cite{Eliezer2003}}. The novelty of SPRAY's laser ray-tracing module lies in the fact that it is coupled to \hl{a} particle-based mes\hl{h}-free SPH method, unlike the de facto mesh-based ray-tracing routine developed by \mbox{\cite{Kaiser2000}} and implemented in state-of-the-art codes like FLASH \mbox{\cite{Fryxell2000}} and HYDRA \mbox{\cite{Marinak2001}}. It also uses equation-of-state (EOS) data generated by MPQEOS \cite{Kemp1998} and opacity data from SNOP \cite{Eidmann1994}. The EOS data is needed for the closure of hydrodynamic equations, and the opacity data, which describes the emission and absorption of radiation, is required for radiation transport calculations. SPRAY is coded in CUDA C/C++, which enables massively parallel execution on NVIDIA GPUs, and is developed on the basis of a well-established nuclear thermal-hydraulics code SOPHIA \cite{Jo2019,Park2020}. Several key modules from SOPHIA, such as the GPU-parallelized nearest neighbor particle search (NNPS) algorithm, along with the predictor-corrector time integration scheme and output file generation routine, were used as the building blocks for SPRAY. The details of numerical methods are presented in \textbf{Section 2}, code benchmark results are shown in \textbf{Section 3}, and conclusions are drawn in \textbf{Section 4}.

\section{Numerical methods}

\subsection{SPH formulations}\label{sec:2.1-1}

The spatial discretization scheme of SPH is composed of two levels of approximation \cite{Liu2008}. The integral representation of a function $f$ can be expressed as
\begin{equation}\label{eq:1}
    f(x)=\int_{-\infty}^\infty{f(x')\delta(x-x')dx'}
\end{equation}
where $\delta(x)$ is the Dirac delta function. The first level of approximation, which is called kernel approximation, approximates the Dirac delta function with a kernel function that has a finite and compact support domain:
\begin{equation}\label{2.1-2}
    f(x)\approx\int_\Omega{f(x')W(x-x',h)dx'}
\end{equation}
where $W(x,h)$ is the aforementioned kernel function, and $h$, which is the characteristic length of the support domain, is called "smoothing length". In general, the kernel function needs to satisfy the following conditions:
\begin{enumerate}
    \item $\lim_{h\rightarrow 0}{W(x,h)}=\delta(x)$ (Delta function-like)
    \item $W(x,h)=W(-x,h)$ (symmetric)
    \item $W(x,h)=0$ for $|x|>\kappa h$, $\kappa$ is a constant (finite support domain)
    \item $W(x,h) \ge 0$ for $\forall x$ (positive)
\end{enumerate}

In SPRAY, Wendland C\textsuperscript{2} kernel \cite{Wendland1995} is used as the de facto kernel function, as it is known to be stable against pairing instabilities \cite{Dehnen2012}. For 3D, the kernel function is defined as
\begin{equation}\label{2.1-3}
    W(x,h)=
    \begin{cases}
        \frac{21}{16\pi h^3}(1-\frac{q}{2})^4(2q+1) & q<2\\
        0 & q\ge 2
    \end{cases}
\end{equation}
where $q=|x|/h$.

The second level of approximation, which is particle approximation, discretizes the volume integral into a sum of neighboring particles:

\begin{equation}\label{2.1-4}
    f(x_a)\approx\sum_b{f(x_b)W(x_a-x_b,h_a)V_b}=\sum_b{f(x_b)W(x_{ab},h_a)\frac{m_b}{\rho_b}}
\end{equation}
where the indices $a$ and $b$ are used to represent each particle, $V$ is volume, $m$ is mass and $\rho$ is mass density. The summation is over all neighboring particles within the support domain. Here, the relation $\rho=m/V$ is used in the second equality, and the abridged notation $x_{ab}=x_a-x_b$ will be used throughout this paper. Also, the index $a$ is added to the smoothing length $h$ to account for the case where the smoothing length varies from particle to particle.

\subsubsection{Momentum conservation}\label{sec:2.1.1}

The fluid equation of motion in Lagrangian frame reads

\begin{equation}\label{2.1.1-5}
    \frac{D\boldsymbol{v}}{Dt}=-\frac{1}{\rho}\nabla P
\end{equation}
where $D/Dt=\left(\partial/\partial t+v\cdot\nabla\right)$ is the material derivative. Here the viscous term present in the Navier-Stokes equation is absent, as the target of simulations is high energy density plasma.

There are multiple ways to formulate SPH equations for first order derivatives. If equation \eqref{2.1-2} is differentiated and integrated by parts, the following relation can be obtained:

\begin{equation}\label{2.1.1-6}
    \frac{d}{dx}f(x)\approx\int_\Omega{f(x')\frac{d}{dx}W(x-x',h)dx'}
\end{equation}

In the process of deriving this equation, the surface integral is assumed to vanish due to the compact nature of the support domain. This form of gradient approximation, despite being simple and easy to implement, suffers from numerical issues and is inferior to more sophisticated formulations.

From vector calculus identities, the following relations \cite{Price2012} can be obtained:

\begin{gather}
    \nabla f=\frac{1}{\phi}\left[\nabla(\phi f)-f\nabla\phi\right]\label{2.1.1-7}\\
    \nabla f=\phi\left[\frac{f}{\phi^2}\nabla\phi+\nabla\left(\frac{f}{\phi}\right)\right] \label{2.1.1-8}
\end{gather}
where $\phi$ is an arbitrary, differentiable scalar quantity. The SPH equivalents of these equations are

\begin{gather}
    \nabla f(x_a)=\sum_b{\frac{m_b}{\rho_b}\frac{\phi_b}{\phi_a}\left[f(x_b)-f(x_a)\right]\nabla W(x_{ab},h_a)} \label{2.1.1-9}\\
    \nabla f(x_a)=\sum_b{\frac{m_b}{\rho_b}\left[\frac{\phi_b}{\phi_a}f(x_a)+\frac{\phi_a}{\phi_b}f(x_b)\right]\nabla W(x_{ab},h_a)}\label{2.1.1-10}
\end{gather}
respectively. It is known that these two formulations form a conjugate pair, provided that the definition of $\phi$ is consistent \cite{Cummins1999}. From preliminary testing involving HEDP, it was evident that formulations with $\phi=1$ (used in \cite{Ritchie2001}) yield more accurate results, compared to other popular choices such as $\phi=\rho$ \cite{Cummins1999}, $\phi=\rho^{1/2}$ \cite{Marri2003} or $\phi=\sqrt{P}/\rho$ \cite{Hernquist1989}. Therefore, the equation of motion in SPRAY follows the form of equation \eqref{2.1.1-10}, and is expressed as:

\begin{equation}\label{2.1.1-11}
    \frac{\Delta\boldsymbol{v}_a}{\Delta t}=\sum_b{\frac{m_b}{\rho_a\rho_b}\left[p_a\nabla W(x_{ab},h_a)+p_b \nabla W(x_{ab},h_b)\right]}
\end{equation}
where the justification of using $\nabla W\left(x_{ab},h_b\right)$ comes from the derivation involving the least action principle and the Euler-Lagrange equations \cite{Price2012}. The time integration of equation \eqref{2.1.1-11}, along with all other explicit time differential equations shown hereafter, is carried out via predictor-corrector scheme. \hl{The time step size at each step is determined base on the Courant-Friedrichs-Lewy (CFL) conditions, and from the momentum conservation equation, one criterion can be defined based on the rate of change of velocity:}
\begin{equation}\label{2.1.1-12}
    \Delta t=\min\left(\alpha \frac{h_a}{|\Delta \boldsymbol{v}_a/\Delta t|}\right)
\end{equation}
\hl{where $\alpha$ is a coefficient in the order of $0.1$.}

\subsubsection{Energy conservation}\label{sec:2.1.2}

In a laser-driven plasma, the laser energy is first absorbed by electrons, which then transfer that thermal energy to ions and eventually reach thermal equilibrium. Therefore, if the laser pulse time scale is much greater than the electron-ion relaxation time, single temperature description $T=T^i=T^e$ (in this paper, all physics related notations are denoted by superscripts as to avoid confusion with particle indices in the subscript) of the plasma is adequate. However, when dealing with laser pulse lengths comparable to or much shorter than the electron-ion relaxation time, it is necessary to consider the non-equilibrium state ($T^i\neq T^e$) in order to properly describe the initial phase of laser-target interaction dynamics as well as the effects of steep gradients. Furthermore, it is necessary to keep track of the radiation energy density separately, rather than assuming that radiation temperature equals electron temperature ($T^r=T^e$), due to radiation and radiative transfer playing significant roles in the laser-plasma interaction dynamics. \hl{In the laser parameter range this code is targeting at, the radiation temperature deviates far from the electron temperature. Because of this, the assumption of equilibration between electron and radiation temperature yields erroneous results that distort the physical phenomena.} Therefore, a single-fluid three-temperature model is used. The energy equations for ion, electron and radiation are expressed as the following:

\begin{gather}
    \frac{Du^i}{Dt}=-\frac{p^i}{\rho}\nabla\cdot \boldsymbol{v}+\frac{1}{\rho}\nabla\cdot k^i\nabla T^i +k^{ei}\left(T^e-T^i\right)\label{2.1.2-12}\\
    \frac{Du^e}{Dt}=-\frac{p^e}{\rho}\nabla\cdot \boldsymbol{v}+\frac{1}{\rho}\nabla\cdot k^e\nabla T^e-k^{ei}\left(T^e-T^i\right)+c\kappa^P(E^r-U^P)+q^{laser}\label{2.1.2-13}\\
    \frac{DE^r}{Dt}=\nabla\cdot\left(\frac{c\lambda}{\rho\kappa^R}\nabla E^r\right)-\rho c\kappa^P(E^r-U^P)\label{2.1.2-14}
\end{gather}
Here, $u^i$ and $u^e$ are specific ion/electron internal energy, $E^r$ is the radiation energy density, $k^i$ and $k^e$ are ion/electron thermal conductivity, $k^{ei}$ is the ion-electron heat exchange coefficient, $\kappa^R$ and $\kappa^P$ are Rosseland and Planck mean opacities, $U^P=\frac{4\sigma_B}{c}T^4$ is Planck’s radiation energy density ($\sigma_B$ : Stefan-Boltzmann constant, $c$ : speed of light), \hl{$\lambda$ is the flux limiter that prevents radiation propagation exceeding the physical limit,} and $q^{laser}$ stands for the specific laser energy deposition rate. The second terms on the right hand side of equation \eqref{2.1.2-12} and \eqref{2.1.2-13} are ion and electron thermal conduction terms, respectively, and the third terms correspond to electron-ion heat exchange. The fourth term on the right hand side of equation \eqref{2.1.2-13} and its counterpart in equation \eqref{2.1.2-14} describe the energy exchange between radiation and electron (via radiation emission and absorption), and the rest of equation \eqref{2.1.2-14} is the radiative diffusion term. In the derivation of radiation transport equations, local thermal equilibrium (LTE) is assumed, and flux-limited diffusion approximation and gray approximations are employed. It is worth noting that although it is not a formidable task to implement multigroup diffusion approximation based radiation transport scheme in SPRAY, the gray approximation was chosen as a trade-off between computational burden and physical accuracy. For the same reason, an alternative approach to directly solve the radiative transfer equation \cite{Pawlik2008} was avoided, which is prohibitively expensive.

As discussed in the previous section, the discretization of derivatives in the energy equations should follow the form of equation \eqref{2.1.1-9} with $\phi=1$ so that the equations form a conjugate pair. For the Laplacian terms, numerous studies have proposed SPH formulations for numerical stability and versatility\cite{Brookshaw1985,Schwaiger2008,Fatehi2011,Biriukov2019}. Here, the following SPH formulation is used\cite{Cleary1999}:

\begin{equation}\label{2.1.2-15}
    \nabla\cdot\left(c\nabla f_a\right)=\sum_b{\frac{m_b}{\rho_b}\frac{4c_ac_b}{c_a+c_b}\left(f_a-f_b\right)\frac{\boldsymbol{x}_{ab}\cdot\nabla W(x_{ab},h_a)}{|\boldsymbol{x}_{ab}|^2}}
\end{equation}

Because of the time scale difference, the radiation transport terms in the governing equations are treated separately via operator splitting method, which will be discussed in \textbf{Section \ref{sec:2.4}}. The remaining hydrodynamics terms are calculated in SPRAY in the following form:

\begin{gather}
    \frac{\Delta u_a^i}{\Delta t}=\frac{p_a^i}{\rho_a}\sum_b{\frac{m_b}{\rho_b}(\boldsymbol{v}_a-\boldsymbol{v}_b)\cdot\nabla W(x_{ab},h_a)}+\sum_b{\frac{4m_b}{\rho_a\rho_b}\frac{k_a^ik_b^i}{k_a^i+k_b^i}(T_a^i-T_b^i)\frac{\boldsymbol{x}_{ab}\cdot\overline{\nabla W}(x_{ab})}{|\boldsymbol{x}_{ab}|^2}}+k_a^{ei}(T_a^e-T_a^i)\label{2.1.2-16}\\
    \frac{\Delta u_a^e}{\Delta t}=\frac{p_a^e}{\rho_a}\sum_b{\frac{m_b}{\rho_b}(\boldsymbol{v}_a-\boldsymbol{v}_b)\cdot\nabla W(x_{ab},h_a)}+\sum_b{\frac{4m_b}{\rho_a\rho_b}\frac{k_a^ek_b^e}{k_a^e+k_b^e}(T_a^e-T_b^e)\frac{\boldsymbol{x}_{ab}\cdot\overline{\nabla W}(x_{ab})}{|\boldsymbol{x}_{ab}|^2}}-k_a^{ei}(T_a^e-T_a^i)+q_a^{laser}\label{2.1.2-17}
\end{gather}
where $\overline{\nabla W}\left(x_{ab}\right)=\frac{1}{2}\left(\nabla W\left(x_{ab},h_a\right)+\nabla W\left(x_{ab},h_b\right)\right)$. It can be shown that, by combining the momentum and energy equations, the total system energy is conserved, minus the laser energy deposition source. First, the ion-electron heat exchange terms (third term on the right hand side of equation \eqref{2.1.2-16} and \eqref{2.1.2-17}) trivially conserve energy, and the thermal conduction terms (second term on the right hand side of equation \eqref{2.1.2-16} and \eqref{2.1.2-17}) also conserve energy, considering the fact that all particles share equal mass, and the terms are symmetric with respect to the particle indices $a$ and $b$. For the remaining terms, consider the following relations:

\begin{equation}\label{2.1.2-18}
    \frac{DE_{total}}{Dt}=\sum_a{m_a\left(\boldsymbol{v}_a\cdot\frac{D\boldsymbol{v}_a}{Dt}+\frac{Du_a^i}{Dt}+\frac{Du_a^e}{Dt}\right)}=-\sum_a{\sum_b{\frac{m_am_b}{\rho_a\rho_b}\left[p_a\boldsymbol{v}_b\cdot \nabla W(x_{ab},h_a)+p_b\boldsymbol{v}_a\cdot\nabla W(x_{ab},h_b)\right]}}
\end{equation}

Equation \eqref{2.1.2-18} is identically zero, due to the antisymmetric gradient of the kernel function $\nabla W(x_{ab},h_a)=-\nabla W(x_{ba},h_b)$. Hence, the total energy is conserved up to the numerical precision of the code.

Here, the following formulae were used to obtain electron/ion thermal conductivities and ion-electron heat exchange coefficient \cite{Ramis2012}:

\begin{gather}
    k^e=\frac{3.23(Z+0.24)}{1+0.24Z}\frac{n^eT^e}{m^e\nu^{ei}}\label{2.1.2-19}\\
    k^i=\frac{k^e}{Z^3}\sqrt{\frac{m^e}{m^i}}\left(\frac{T^i}{T^e}\right)^{5/2}\label{2.1.2-20}\\
    k^{ei}=\frac{3m^eZ\nu^{ei}}{(m^i)^2}\label{2.1.2-21}
\end{gather}
where $Z$ is the average ionization level, $\nu^{ei}$ is the electron-ion collision frequency, and temperature is in energy unit ($eV$). In order to correctly model the initial cold, solid state of targets, SPRAY incorporates the electron-phonon model \cite{Eidmann2000} to calculate the collision frequency. The classical Spitzer-H\"arm collision frequency, which is appropriate for hot plasma regime, is interpolated with the electron-phonon collision frequency, and a physical cutoff is introduced to restrict the mean free path relative to the average interatomic distance \cite{Eidmann2000}:

\begin{gather}
    \nu^{ei}_{SH}=\frac{4\sqrt{2\pi}}{3}\frac{n^ie^4}{\sqrt{m^eT^e}}\ln{\Lambda}\\
    \nu^{ei}_{EP}\approx2k_s\frac{e^2T^i}{\hbar^2v^F}\\
    \nu^{ei}=\min\left(\frac{\nu^{ei}_{SH}\nu^{ei}_{EP}}{\nu^{ei}_{SH}+\nu^{ei}_{EP}},r_0\right)
\end{gather}
where $\ln{\Lambda}$ is the Coulomb logarithm, $k_s$ is a numerical constant used as a correction factor based on experimental measurements \cite{Eidmann2000}, $v^F=\hbar(3\pi^2n^e)^{1/3}/m^e$ is the Fermi velocity, $\hbar$ is the Planck's constant, and $r_0$ is the average interatomic distance. $SH$ and $EP$ are abbreviations for Spitzer-H\"arm and electron-phonon models respectively.

In case\hl{s} where even the heat transport time scale is too short for a tractable simulation, an implicit heat transport routine \hl{should be used to evaluate equation \mbox{\eqref{2.1.2-16}} and \mbox{\eqref{2.1.2-17}} in lieu of explicit time integration of said equations with the time step size controlled by CFL conditions. For this reason, an implicit solver} is implemented to \hl{allow for a larger} simulation time step. The thermal conduction terms and the ion-electron relaxation terms from equation \mbox{\eqref{2.1.2-16}} and \mbox{\eqref{2.1.2-17}} discretized in backward Euler scheme \hl{take} the following form:

\begin{gather}
    \frac{u_a^{i,N+1}-u_a^{i,N}}{\Delta t}=\sum_b{\frac{4m_b}{\rho_a\rho_b}\frac{k_a^ik_b^i}{k_a^i+k_b^i}(T_a^{i,N+1}-T_b^{i,N+1})\frac{\boldsymbol{x}_{ab}\cdot\overline{\nabla W}(x_{ab})}{|\boldsymbol{x}_{ab}|^2}}+k_a^{ei}(T_a^{e,N+1}-T_a^{i,N+1})\label{2.1.2-25}\\
    \frac{u_a^{e,N+1}-u_a^{e,N}}{\Delta t}=\sum_b{\frac{4m_b}{\rho_a\rho_b}\frac{k_a^ek_b^e}{k_a^e+k_b^e}(T_a^{e,N+1}-T_b^{e,N+1})\frac{\boldsymbol{x}_{ab}\cdot\overline{\nabla W}(x_{ab})}{|\boldsymbol{x}_{ab}|^2}}-k_a^{ei}(T_a^{e,N+1}-T_a^{i,N+1})\label{2.1.2-26}
\end{gather}
where the superscripts $N$ and $N+1$ refer to the time step index. By extrapolating the temperature as $T^{N+1}=T^N+$ $\frac{\partial T}{\partial u}(u^{N+1}-u^N)$, equation \mbox{\eqref{2.1.2-25}} and \mbox{\eqref{2.1.2-26}} can be written as:

\begin{gather}
    \Lambda_a^iu_a^{i,N+1}+\sum_b{\chi_{ab}^iu_b^{i,N+1}}-\Gamma_a^{e} u_a^{e,N+1} = b_a^i+c_a\label{2.1.2-27}\\
    \Lambda_a^eu_a^{e,N+1}+\sum_b{\chi_{ab}^eu_b^{e,N+1}}+\Gamma_a^{i} u_a^{i,N+1} = b_a^e-c_a\label{2.1.2-28}
\end{gather}
where
\begin{gather}
    \Lambda_a^s=1-\Delta t\left(\frac{\partial T^s}{\partial u^s}\right)_a\sum_b{\frac{4m_b}{\rho_a\rho_b}\frac{k_a^sk_b^s}{k_a^s+k_b^s}\frac{\boldsymbol{x}_{ab}\cdot\overline{\nabla W}(x_{ab})}{|\boldsymbol{x}_{ab}|^2}}+\Delta t k^{ei}_a\left(\frac{\partial T^s}{\partial u^s}\right)_a\\
    \chi_a^s=\Delta t\sum_b{\frac{4m_b}{\rho_a\rho_b}\frac{k_a^sk_b^s}{k_a^s+k_b^s}\frac{\boldsymbol{x}_{ab}\cdot\overline{\nabla W}(x_{ab})}{|\boldsymbol{x}_{ab}|^2}\frac{\partial T^s_b}{\partial u^s_b}u_b^{s,N+1}}\\
    \Gamma_a^s=\Delta t k^{ei}_a\left(\frac{\partial T^s}{\partial u^{s}}\right)_au^{s,N+1}_a\\
    b_a^s = u^{s,N}_a+\Delta t\sum_b{\frac{4m_b}{\rho_a\rho_b}\frac{k_a^sk_b^s}{k_a^s+k_b^s}\frac{\boldsymbol{x}_{ab}\cdot\overline{\nabla W}(x_{ab})}{|\boldsymbol{x}_{ab}|^2}\left[T_a^{s,N}-T_b^{s,N}-\left(\frac{\partial T^s}{\partial u^s}\right)_au_a^{s,N}+\left(\frac{\partial T^s}{\partial u^s}\right)_bu_b^{s,N}\right]}\\
    c_a = \Delta t k^{ei}\left[T_a^{e,N}-T_a^{i,N}-\left(\frac{\partial T^e}{\partial u^e}\right)_au_a^{e,N}+\left(\frac{\partial T^i}{\partial u^i}\right)_au_a^{i,N}\right]
\end{gather}
The superscript $s$ refers to the species ($e$ for electron, $i$ for ion).

When there are $n$ number of particles, the equations \mbox{\eqref{2.1.2-27}} and \mbox{\eqref{2.1.2-28}} form a linear system $Ax=b$ with $A$ being a $2n\times2n$ matrix, and $x$ and $b$ being $2n\times1$ vectors. Considering that the number of neighbors for each particle is relatively constant, the sparsity of the matrix $A$ increases as the number of particles increases. For a modest two dimensional test case with 200,000 particles, the sparsity was measured to be greater than 99.9\%. Due to the restriction of GPU memory, it is unrealistic to directly solve the linear system with LU factorization or QR factorization methods. Instead, Krylov subspace\hl{-}based iterative method\hl{s are} used. In the initial testing on CPU\hl{s}, \hl{the} bi-conjugate gradient stabilized (BiCGSTAB) method \mbox{\cite{vanDerVorst1992}} with incomplete LU preconditioning yielded the highest performance compared to other popular methods such as \hl{the} generalized minimal residual (GMRES) method \mbox{\cite{Saad1986}}. Therefore, \hl{the} GPU\hl{-}accelerated incomplete LU preconditioner and BiCGSTAB solver are implemented using cuBLAS and cuSPARSE libraries, and are coupled to \hl{the} SPRAY code.

\subsubsection{Mass conservation}\label{sec:2.1.3}
    
 In SPH, mass is inherently conserved, since the mass of each particle is kept constant. However, as the particles move and their volumes evolve, their densities are updated. There are two ways to estimate the density of each particle in SPH. The first way is to take the kernel weighted average of neighboring particles’ densities:

 \begin{equation}\label{2.1.3-34}
     \rho_a=\sum_b{\frac{m_b}{\rho_b}\rho_b W(x_{ab},h_a)}=\sum_b{m_b W(x_{ab},h_a)}
 \end{equation}

 Despite its simplicity, this approach has a major drawback in the context of laser-plasma interaction simulation. On the free surface boundary of the plasma, there inevitably is a deficiency of particles within the support domain of the kernel function \cite{Monaghan1994}. As a result, fewer neighboring particles participate in the summation in equation \eqref{2.1.3-34}, thereby causing density underestimation. Numerous studies have proposed various solutions to address this issue, which will be discussed in detail in \textbf{Section \ref{sec:2.2}}. However, none of these solutions were effective and versatile enough to be applied to high-intensity laser-plasma interaction simulations.

The other way of estimating density is based on the continuity equation:

\begin{equation}\label{2.1.3-35}
    \frac{D\rho}{Dt}=-\rho\nabla\cdot\boldsymbol{v}
\end{equation}

Keeping with the conjugate pairs, equation \eqref{2.1.3-35} is discretized in the form of equation \eqref{2.1.1-9}:

\begin{equation}\label{2.1.3-36}
    \frac{\Delta \rho_a}{\Delta t}=\rho_a\sum_b{\frac{m_b}{\rho_b}(\boldsymbol{v}_a-\boldsymbol{v}_b)\cdot\nabla W(x_{ab},h_a)}
\end{equation}
\hl{With the evaluation of this equation, another criterion for the CFL condition can be defined:}
\begin{equation}\label{2.1.3-39}
     \Delta t=\min\left(\beta \frac{\rho_a}{|\Delta \rho_a/\Delta t|}\right)
\end{equation}
\hl{where $\beta$ is a coefficient in the order of 0.1.}

In addition, the kernel gradient correction (KGC) filter \cite{Shao2012} is applied to the continuity equation to improve the accuracy of density estimation, especially near the free surface:

\begin{gather}
    \left(\nabla W(x_{ab},h_a)\right)^{corrected}=\boldsymbol{L}\nabla W(x_{ab},h_a)\label{2.1.3-37}\\
    \boldsymbol{L}= \left(\sum_b{\frac{m_b}{\rho_b}
    \begin{pmatrix}
        x_{ba}\frac{\partial W}{\partial x_a} & y_{ba}\frac{\partial W}{\partial x_a} & z_{ba}\frac{\partial W}{\partial x_a} \\
        x_{ba}\frac{\partial W}{\partial y_a} & y_{ba}\frac{\partial W}{\partial y_a} & z_{ba}\frac{\partial W}{\partial y_a} \\
        x_{ba}\frac{\partial W}{\partial z_a} & y_{ba}\frac{\partial W}{\partial z_a} & z_{ba}\frac{\partial W}{\partial z_a}
    \end{pmatrix}
    }\right)^{-1}\label{2.1.3-38}
\end{gather}

\subsection{Energy conserving numerical dissipation schemes}\label{sec:2.3}

Due to the presence of shock waves, numerical dissipation schemes in the form of artificial viscosity are essential. There have been a number of different numerical dissipation schemes proposed, each of them tuned to the specific problems that are being solved \cite{Monaghan1997}. In SPRAY, the numerical dissipation scheme proposed by Price \cite{Price2012} is implemented, which not only conserves energy exactly, but also controls the degree of numerical treatment with a dissipation switch.

\begin{figure}[htbp]
    \centering
    \includegraphics[width=0.4\linewidth]{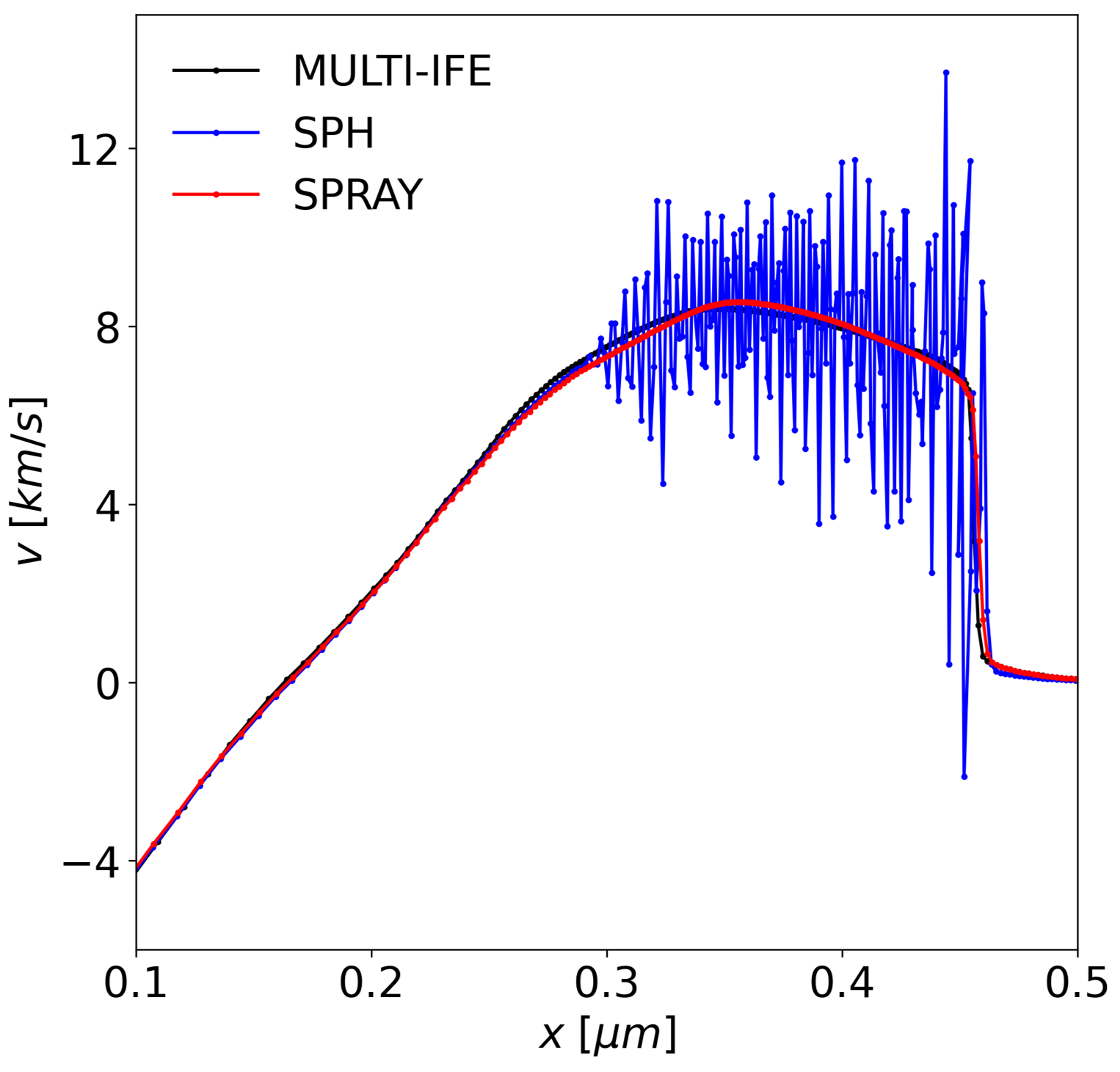}
    \caption{\textmd{Velocity profile of laser irradiation simulations with and without the energy conserving numerical dissipation scheme applied. The sharp gradient at $x=0.45\ \mu m$ is the shock front, and the unphysical oscillations propagate inward from the shock front into the compression region. The reference profile is calculated by MULTI-IFE \cite{Ramis2016}. The SPRAY and MULTI-IFE profiles are overlapped due to yielding almost identical results.}}
    \label{fig:4}
\end{figure}

The numerical dissipation terms are included in the momentum equation and the ion energy equation, since ion carries the bulk of the momentum and thus is responsible for the viscous effects.

\begin{gather}
    \left(\frac{D\boldsymbol{v}_a}{Dt}\right)_{diss}=\sum_b{\frac{m_b}{2}\left[\frac{v_{sig,a}}{\rho_a}\nabla W(x_{ab},h_a)+\frac{v_{sig,b}}{\rho_b}\nabla W(x_{ab},h_b)\right](\boldsymbol{v}_{ab}\cdot\hat{\boldsymbol{x}}_{ab})}\label{2.3-42}\\
    \left(\frac{Du_a^i}{Dt}\right)_{diss}=-\frac{1}{\rho_a}\sum_b{\frac{m_bv_{sig,a}}{2}(\boldsymbol{v}_{ab}\cdot\hat{\boldsymbol{x}}_{ab})^2\hat{\boldsymbol{x}_{ab}}\cdot \nabla W(x_{ab},h_a)}\label{2.3-43}
\end{gather}
where $\hat{\boldsymbol{x}}_{ab}=\boldsymbol{x}_{ab}/|\boldsymbol{x}_{ab}|$, and $v_{sig,a}$ is signal velocity defined as:

\begin{equation}\label{2.3-44}
    v_{sig,a}=\alpha_ac_a^s+\beta|\boldsymbol{v}_{ab}\cdot\hat{\boldsymbol{x}}_{ab}|
\end{equation}
Here, $c^s$ is the ion sound speed, $\beta=2$ controls and prevents particle interpenetration, and $\alpha$ is the aforementioned dissipation switch, which is automatically adjusted according to

\begin{equation}\label{2.3-45}
    \frac{d\alpha_a}{dt}=\max{(-\nabla\cdot\boldsymbol{v},0)}-\frac{\alpha_a-\alpha_{min}}{\tau_a}
\end{equation}
where $\tau_a=h_a/(\sigma_{decay}v_{sig,a})$ and $\alpha_{min}$ is the minimum value for the switch. By implementing this switch, the dissipation effect is localized to the shock fronts, and decays in magnitude away from them. \hl{The appropriate time step size can be determined by introducing yet another CFL condition criterion:}
\begin{equation}\label{2.3-47}
    \Delta t=\min\left(\gamma\frac{h_a}{\max(c^s_{a},v_{sig,a})}\right)
\end{equation}
\hl{where $\gamma$ is a coefficient in the order of 0.1.} The robustness of this numerical dissipation scheme is illustrated in \textbf{Figure \ref{fig:4}}. The SPRAY profile is barely distinguishable from the reference profile, due to yielding almost identical results. The problem setup is identical to that of \textbf{Figure \ref{fig:3}}. The velocity profile smoothness is much improved when the numerical dissipation is applied, compared to the default SPH scheme with no additional treatments. The unphysical oscillations, which are formed due to the sharp gradient created at the shock front, propagate inward toward the compression region to the right, which compromises the accuracy and stability of evaluating the hydrodynamics governing equations.

The energy conserving property of this numerical dissipation scheme can be shown in a straightforward manner. In a similar way with equation \eqref{2.1.2-18},

\begin{multline}\label{2.3-46}
    \left(\frac{DE_{total}}{Dt}\right)_{diss}=\sum_a{m_a\left[\boldsymbol{v}_a\cdot\left(\frac{D\boldsymbol{v}_a}{Dt}\right)_{diss}+\left(\frac{Du_a^i}{Dt}\right)_{diss}\right]}\\
    =\sum_a{\sum_b{\frac{m_am_b}{2}\left[\frac{\boldsymbol{v}_av_{sig,b}\cdot\nabla W(x_{ab},h_b)}{\rho_b}+\frac{\boldsymbol{v}_bv_{sig,a}\cdot\nabla W(x_{ab},h_a)}{\rho_a}\right]}}
\end{multline}
By the same logic, the antisymmetric property of the kernel function gradient ensures that equation \eqref{2.3-46} is identically zero.

\subsection{Radiation transport}\label{sec:2.4}

\begin{figure}[htbp]
    \centering
    \includegraphics[width=0.6\linewidth]{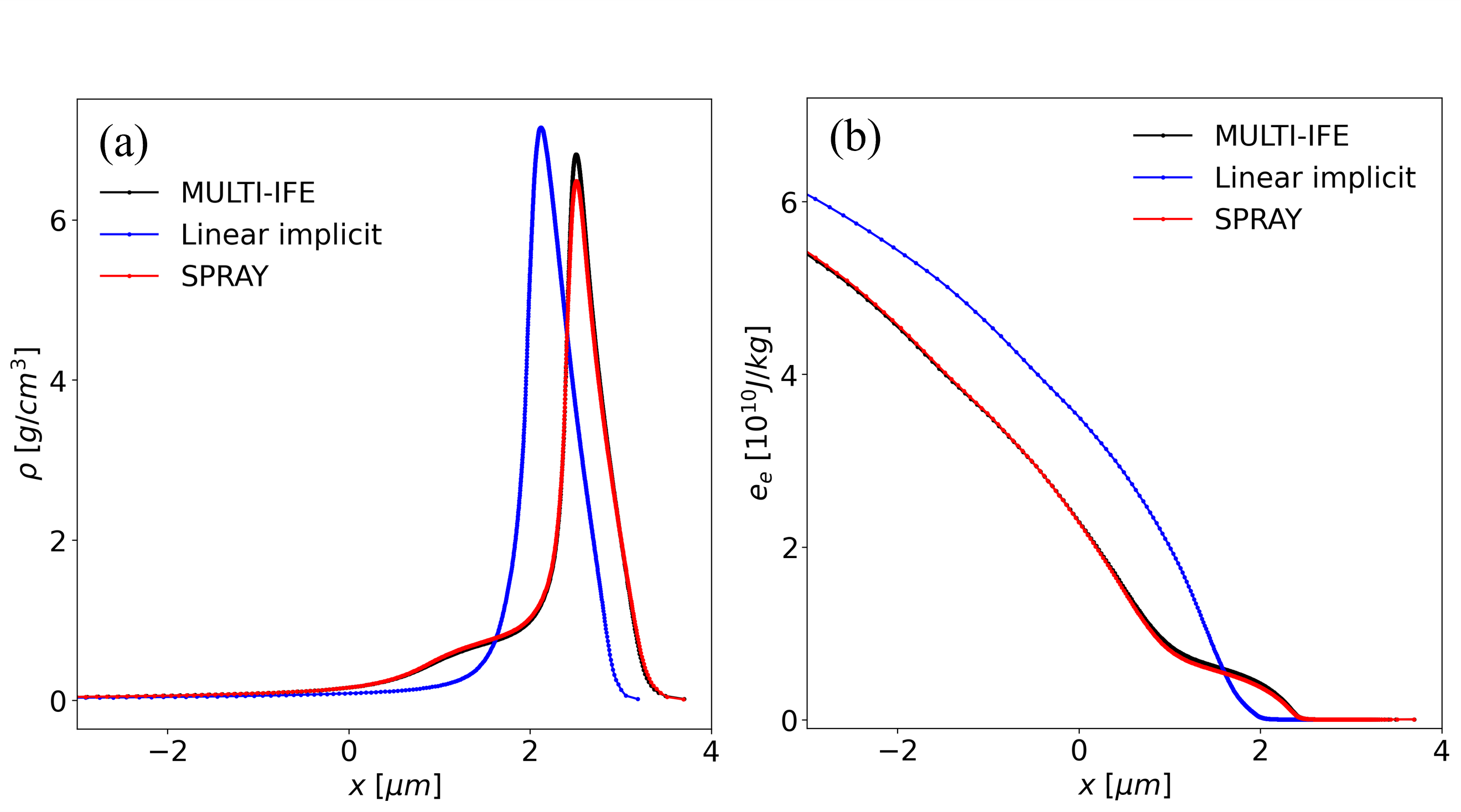}
    \caption{\textmd{Improved radiation transport calculation accuracy with explicit quartic scheme. (a) and (b) are density and specific electron internal energy profiles, respectively. Blue line represents the calculation results when solving the implicit formulations with $U^{P,n+1}=U^{P,n}+\frac{\partial U^P}{\partial T^e}\frac{\partial T^e}{\partial u^e}\Delta u^e$, and the red line represents the results for explicit handling of the quartic dependence of the Planck energy density on electron temperature, i.e. $U^{P,n+1}\approx \frac{4\sigma_B}{c}\left(T^{e,n}+\frac{\partial T^e}{\partial u^e}\Delta u^{e,n} \right)^4$. The reference profiles are calculated by MULTI-IFE \cite{Ramis2016}. The SPRAY and MULTI-IFE profiles are overlapped due to yielding almost identical results.}}
    \label{fig:5}
\end{figure}

In SPRAY, radiation transport is solved separately due to the radiative transfer time scale being much shorter than the hydrodynamic time scale. In order to efficiently solve the radiation transport equations, a backward Euler based implicit scheme with Jacobi iteration \cite{Whitehouse2005} is used. The SPH formulation of the governing equations are as follows:
\begin{gather}
    \frac{\Delta E_a^r}{\Delta t}=\sum_b{\frac{m_b}{\rho_b}\frac{4\mu_a\mu_b}{\mu_a+\mu_b}(E_a^r-E_b^r)\frac{\boldsymbol{x}_{ab}\cdot\nabla W(x_{ab},h_a)}{|\boldsymbol{x}_{ab}|^2}}+\rho_ac\kappa_a^P(U_a^P-E_a^r)\label{2.4-47}\\
    \frac{\Delta u_a^e}{\Delta t}=c\kappa^P_a(E_a^r-U^P_a)\label{2.4-48}
\end{gather}
where $\mu=\frac{c\lambda}{\rho\kappa^R}$, $\lambda$ being the flux limiter that prevents radiation propagation exceeding the physical limit \hl{in optically thin channels that might arise during the evolution of the system}. In this study, the Larsen flux limiter is used $\lambda=1/(3^k+R^k)^{1/k}$ where $k=2$ and $R\equiv |\nabla E|/(\rho \kappa^R E) $\mbox{\cite{Morel2000}}. \hl{This flux limiter takes the value of $1/3$ in highly optically thick regime, thus restoring the original equation, and approaches zero in the optically thin region.} The equations above can be discretized into implicit form with backward Euler scheme where the superscripts denote the time index:
\begin{gather}
    E_a^{r,n+1}=E_a^{r,n}+dt\Lambda_aE_a^{r,n+1}+dt\Gamma_a+dt\rho_ac\kappa_a^P(U_a^{P,n+1}-E_a^{r,n+1})\label{2.4-49}\\
    u_a^{e,n+1}=u_a^{e,n}+dtc\kappa_a^P(E_a^{r,n+1}-U_a^{P,n+1})\label{2.4-50}\\
    \Lambda_a\equiv\sum_b{\frac{m_b}{\rho_b}\frac{4\mu_a\mu_b}{\mu_a+\mu_b}\frac{\boldsymbol{x}_{ab}\cdot\nabla W(x_{ab},h_a)}{|\boldsymbol{x}_{ab}|^2}}\\
    \Gamma_a\equiv\sum_b{\frac{m_b}{\rho_b}\frac{4\mu_a\mu_b}{\mu_a+\mu_b}(-E_b^{r,n+1})\frac{\boldsymbol{x}_{ab}\cdot\nabla W(x_{ab},h_a)}{|\boldsymbol{x}_{ab}|^2}}
\end{gather}
Here, the extrapolation of Planck energy density is handled in the following manner:
\begin{equation}\label{2.4-53}
    U^{P,n+1}=\frac{4\sigma_B}{c}(T_a^{e,n+1})^4\approx a\left(T_a^{e,n}+\frac{\partial T^e}{\partial u^e}(u_a^{e,n+1}-u_a^{e,n})\right)^4
\end{equation}
where $a=\frac{4\sigma_B}{c}$ is the radiation constant and $\sigma_B$ is the Stefan-Boltzmann constant. The nonlinear quartic dependence of the Planck energy density on electron temperature is \hl{retained \mbox{\cite{Jiang2021,Menon2022}}} rather than by a common, straightforward linearization approach involving the gradient $\partial U^P/\partial T^e$, i.e. $U^{P,n+1}\approx U^{P,n}+\frac{\partial U^P}{\partial T^e}\frac{\partial T^e}{\partial u^e}\Delta u^e$ \mbox{\cite{Whitehouse2005,Bassett2021A,Bassett2021B}}. This is because, although the latter form avoids the arduous task of solving quartic equations at every time step, serious over/underestimation in the extrapolation process can be caused as the hydrodynamical time step is taken (see \textbf{Figure \ref{fig:5}}). For this reason, recently suggested approaches to accelerate the radiation transport routine could not be used, as those methods are based on the assumption that the linearization of the Planck function is acceptable \mbox{\cite{Bassett2021A,Bassett2021B}}. However, the convergence speed of our approach was within an acceptable range for laser intensities \hl{up to} $10^4$ TW/cm\textsuperscript{2}, beyond which \hl{more sophisticated} relativistic treatment\hl{s are required}. Then, the equations can be rewritten in the following forms:
\begin{gather}
    (1-dt\Lambda_a+dt\rho_ac\kappa_a^P)E_a^{r,n+1}=E_a^{r,n}+dt\Gamma_a+dt\rho_ac\kappa_a^Pa\chi^4\label{2.4-54}\\
    E_a^{r,n+1}=\frac{1}{dtc\kappa_a^P}(u_a^{e,n+1}-u_a^{e,n})+a\chi_a^4\label{2.4-55}\\
    \chi_a\equiv T_a^{e,n}+\frac{\partial T^e}{\partial u^e}(u_a^{e,n+1}-u_a^{e,n})
\end{gather}

By substituting equation \eqref{2.4-55} into \eqref{2.4-54}, a quartic equation of $\chi_a$ can be obtained by rearranging the terms:
\begin{equation}\label{2.4-57}
    (1-dt\Lambda_a)a\chi_a^4+\frac{1-dt\Lambda_a+dt\rho_ac\kappa_a^P}{dtc\kappa_a^P}\left(\frac{\partial T^e}{\partial u^e}\right)^{-1}\chi_a\\
    -\left[E_a^{r,n}+dt\Gamma_a+\frac{1-dt\Lambda_a+dt\rho_ac\kappa_a^P}{dtc\kappa_a^P}\left(\frac{\partial T^e}{\partial u^e}\right)^{-1}T_a^{e,n}\right]=0
\end{equation}

By solving equation \eqref{2.4-57} for $\chi_a$, $E_a^{r,n+1}$ and $u_a^{e,n+1}$ can be obtained from equation \eqref{2.4-54}. When solving the quartic equation, the root that yields a real value of $u_a^{e,n+1}$ closest to $u_a^{e,n}$ is taken out of four possible roots. Because the cubic and quadratic terms are absent, the four roots could be found directly for the depressed quartic equation $x^4+bx+c=0$. Defining $p=-c$ and $q=-b^2/8$, the four roots are:
\begin{equation}
    x=\frac{1}{2}\left(\pm\sqrt{2y}\pm\sqrt{-2y+\frac{2b}{\sqrt{2y}}}\right)
\end{equation}
where
\begin{gather}
    y=w-\frac{p}{3w}\\
    w=\sqrt[3]{-\frac{q}{2}+\sqrt{\frac{q^2}{4}+\frac{p^3}{27}}}
\end{gather}

This Jacobi iteration is repeated until $E_a^{r,n+1}$ and $u_a^{r,n+1}$ converge. \hl{At each iteration, the coefficients of equation \mbox{\eqref{2.4-57}} are updated, and the quartic equation is solved for each particle.} The convergence criterion is taken as the relative error of the difference between the two sides of equation \eqref{2.4-49}.

The method of implicitly solving radiation transport described in this section can be easily \hl{extended to a} multi-group diffusion model. However, the convergence of the Jacobi iteration would be severely slowed, and currently there is no efficient method to solve multi-group transport under the SPH formulation while retaining the quartic nonlinearity of the Planck function. Further improvement in this approach would be imperative for a more complete description of radiation transport.

\subsection{Free surface boundary}\label{sec:2.2}

\hl{SPH was originally developed to study astrophysical plasmas, where boundaries are neglected. The lack of clear delineations of surfaces like in mesh-based approaches makes boundaries a non-trivial issue} \cite{Lind2020}. In the formulation of SPH equations, it is assumed that sufficient particles are present within the kernel support domain (refer to equation \eqref{2.1-4}). However, on or near the boundary, the truncation of the SPH kernel, i.e. particle deficiency, leads to inaccurate interpolation, thus compromising the accuracy of the calculation. There have been numerous studies proposing solutions to this issue, including the Shepard filter \cite{Shepard1968}, kernel gradient correction (KGC) \cite{Shao2012}, corrective smoothed particle method (CSPM) \cite{Chen1999}, and finite particle method (FPM) \cite{Liu2005}. Although these methods are shown to be robust and versatile in many applications, they are not suitable for freely expanding surfaces where the variation of density is very dramatic and the gradient on the boundary surface could be very steep.

\begin{figure}[htbp]
    \centering
    \includegraphics[width=0.6\linewidth]{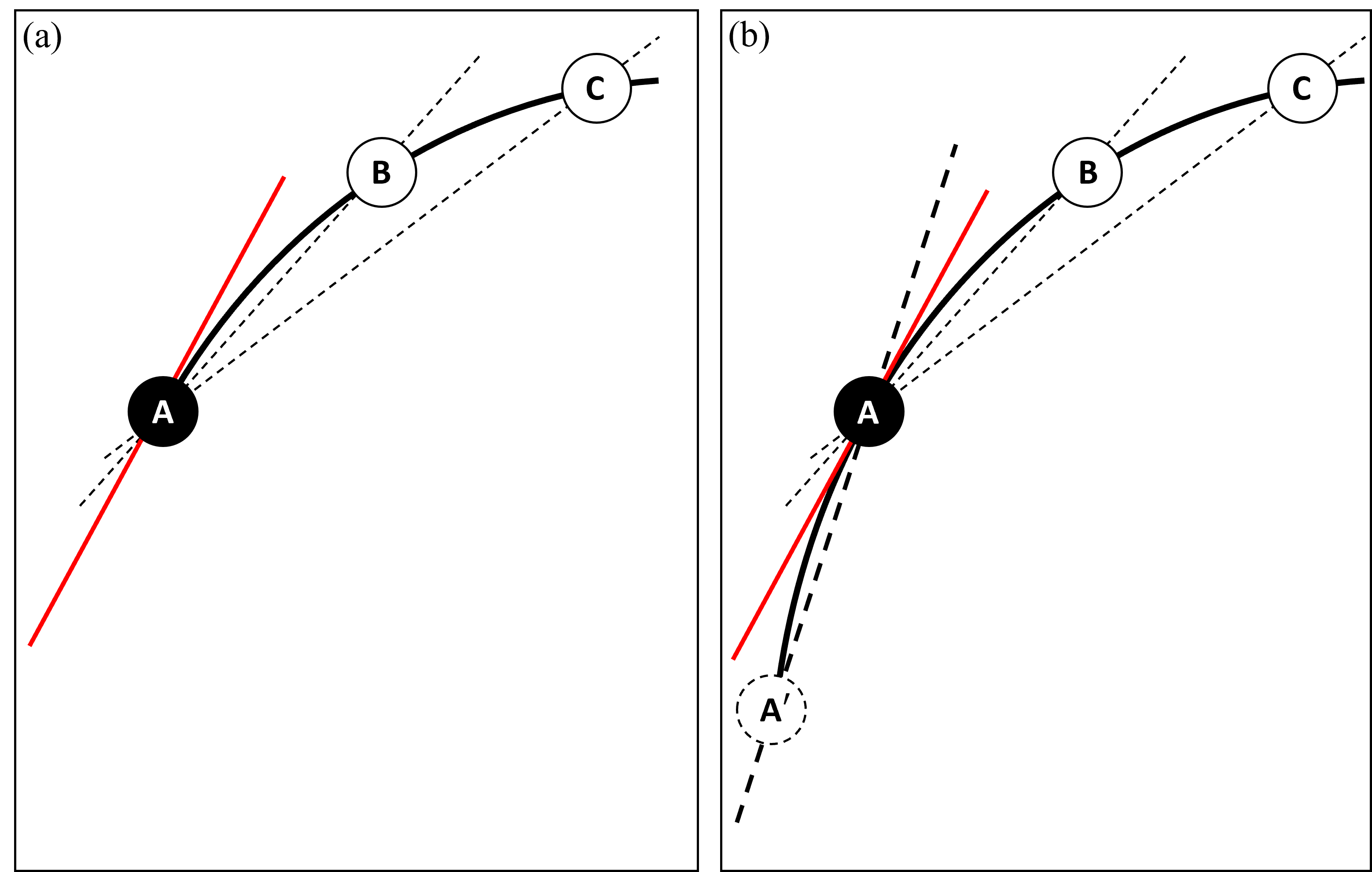}
    \caption{\textmd{Illustration of the issue in estimating the gradient of a physical profile near the free surface boundary. For the outermost boundary particle A in (a), the true gradient (red line) is greater than the approximated gradient from neighbor particles B and C (dashed line). Hence, a virtual, mirrored particle ($A^\prime$) needs to be present to serve as a stencil and to recover the true gradient (see (b)).}}
    \label{fig:2}
\end{figure}

When simulating laser irradiation of a target, the target surface undergoes ablation, and a rapidly expanding coronal plasma is formed. As a result, a steep gradient is formed in the physical profiles (density, pressure, velocity, etc.) where the difference in density \hl{over a few dozen SPH particles apart} can be in the order of $10^3-10^4$. Moreover, the gradient becomes steeper towards the edge, because the second derivative of the density profile in the vicinity of the free surface is negative (see \textbf{Figure \ref{fig:2}(a)}). This poses unique challenges in boundary treatment, since the SPH approximation would result in underestimation of the gradient due to lack of sufficient information. In other words, the conventional scheme fails to recover the true gradient since all of the neighboring particles are on the flatter side of the gradient.

There have been attempts to address similar problems. Reinhardt et al. \cite{Reinhardt2017} suggested using a correction factor that is determined a priori by numerical fitting of the sample data. The essence of this scheme is to estimate the true gradient based on the corrected volume of each particle, which is determined based on an imbalance factor that holds the information of the particle distribution. The shortcomings of this approach not only include the issue where the numerical fitting needs to be performed every time a different kernel function is chosen, but also lie in the fact that this is essentially an a posteriori fix with limited physical justifications. Therefore, the utility of this scheme is highly dependent on the characteristics of the problem at hand, and was shown to be limited in accuracy in our testing. On the other hand, Ruiz-Bonilla et al. \cite{Ruiz-Bonilla2022} proposed estimating density from the equation-of-state using estimated pressure and temperature. The key idea was to estimate the pressure and temperature of the boundary particles using nearby particles, and then estimate the density corresponding to that state of matter. Although this approach is physically more sound, it would yield accurate and reliable results only when the pressure and temperature profiles vary smoothly with respect to the smoothing length scale. Unfortunately, this is not the case in typical high intensity laser-target interaction problems.

In SPRAY, a novel scheme is developed and implemented to handle this freely expanding surface boundary. In order to solve the hydrodynamics governing equations, the divergence of velocity and the gradient of pressure need to be accurately estimated (please refer to equations \mbox{\eqref{2.1.1-5}}, \mbox{\eqref{2.1.2-12}-\eqref{2.1.2-13}}, and \mbox{\eqref{2.1.3-35}}). The scheme proposed in this study improves the accuracy of these gradient estimation\hl{s} with a combination of a free surface tracking scheme and a particle mirroring method (see \mbox{\textbf{Figure \ref{fig:2}(b)}}).

First, the position of the free surface needs to be located. A free surface tracking scheme, which solves the equation of motion of this boundary, is employed to keep track of this information. In a staggered cell configuration, the velocity of the outermost surface due to the pressure gradient force can be calculated by the following simple equation:
\begin{equation}
    \frac{\Delta v_i}{\Delta t}=\frac{2(P_{i-1}-P_i)}{m_{i-1}+m_i} \rightarrow \frac{\Delta v_1}{\Delta t}=-\frac{2P_1}{m_1}
\end{equation}
where the subscript index refers to the cell/interface index, and index of 1 corresponds to the outermost cell/interface.

For multidimensional simulation with arbitrary particle distribution, the free surface can be tracked by the following algorithm. First, the boundary particles are identified using the cover vector method \mbox{\cite{Barecasco2013}}. The relative position vector with respect to neighbor particles $c_a=\sum_b{r_{ab}/{|r_{ab}|}}$, defined as the cover vector, is computed for each particle during a SPH loop. This cover vector serves as a rough estimation of the surface normal. Then, the presence of neighbor particles within the angular vicinity of the cover vector, defined as the scan cone, is examined. If there are no particles within a prescribed scan angle, then that particle is identified as a boundary particle. Then, the position of a point on the free surface is located using both the surface normal unit vector and the particle volume. For the surface normal vector, the cover vector is corrected with the particle velocity to determine the surface normal unit vector more reliably ($\hat{n}_a=\alpha \hat{c}_a + (1-\alpha) \hat{v}_a$, $0<\alpha<1$). For the particle volume, the relation between the smoothing length and particle volume is used, i.e. $h_a=h_{fact}(m_a/\rho_a)^{1/d}$ where $d$ is the dimension, and $h_{fact}$ is a constant. Using both the surface normal unit vector and the particle volume, a point on the free surface can be located. The relative velocity $\boldsymbol{\tilde{v}}$ of that point on the free surface with respect to the boundary particle can be estimated with the conventional SPH momentum equation (zero pressure on the free surface is assumed).
\begin{equation}
    \frac{\Delta\boldsymbol{\tilde{v}}_{free-surface,a}}{\Delta t}=\frac{m_a}{\rho_a^2}p_a\nabla W(x_{free-surface,a},h_a)
\end{equation}
This equation is consistent with Equation \mbox{\eqref{2.1.1-11}} where the neighboring particle has zero pressure.

When the location of the free surface is identified, the particle mirroring method is used to improve the accuracy of the gradient estimation. The gist of this particle mirroring is to create a numerical stencil that contributes to the gradient estimation by compensating for the underestimation of conventional approaches. Recall that the gradients that need to be evaluated are those of the velocity and pressure profile. Hence, the information that needs to be defined includes the (extrapolated) velocity, pressure, and the position of the stencil particle. These values are determined via linear extrapolation with respect to the free surface. For instance, the velocity of the mirrored particle is defined as:
\begin{equation}
    v_{mirrored}=2\cdot v_{free-surface}-v_{outermost}
\end{equation}
where $v_{outermost}$ is the velocity of the boundary particle identified in the previous step, and $v_{free-surface}$ is obtained from the free surface tracking scheme. The same method is used for both the position and pressure. The position of the free surface is also given by the free surface \hl{tracking} scheme, and the pressure of the free surface is assumed to be zero, which is a physically sound assumption.

\begin{figure}[htbp]
    \centering
    \includegraphics[width=0.7\linewidth]{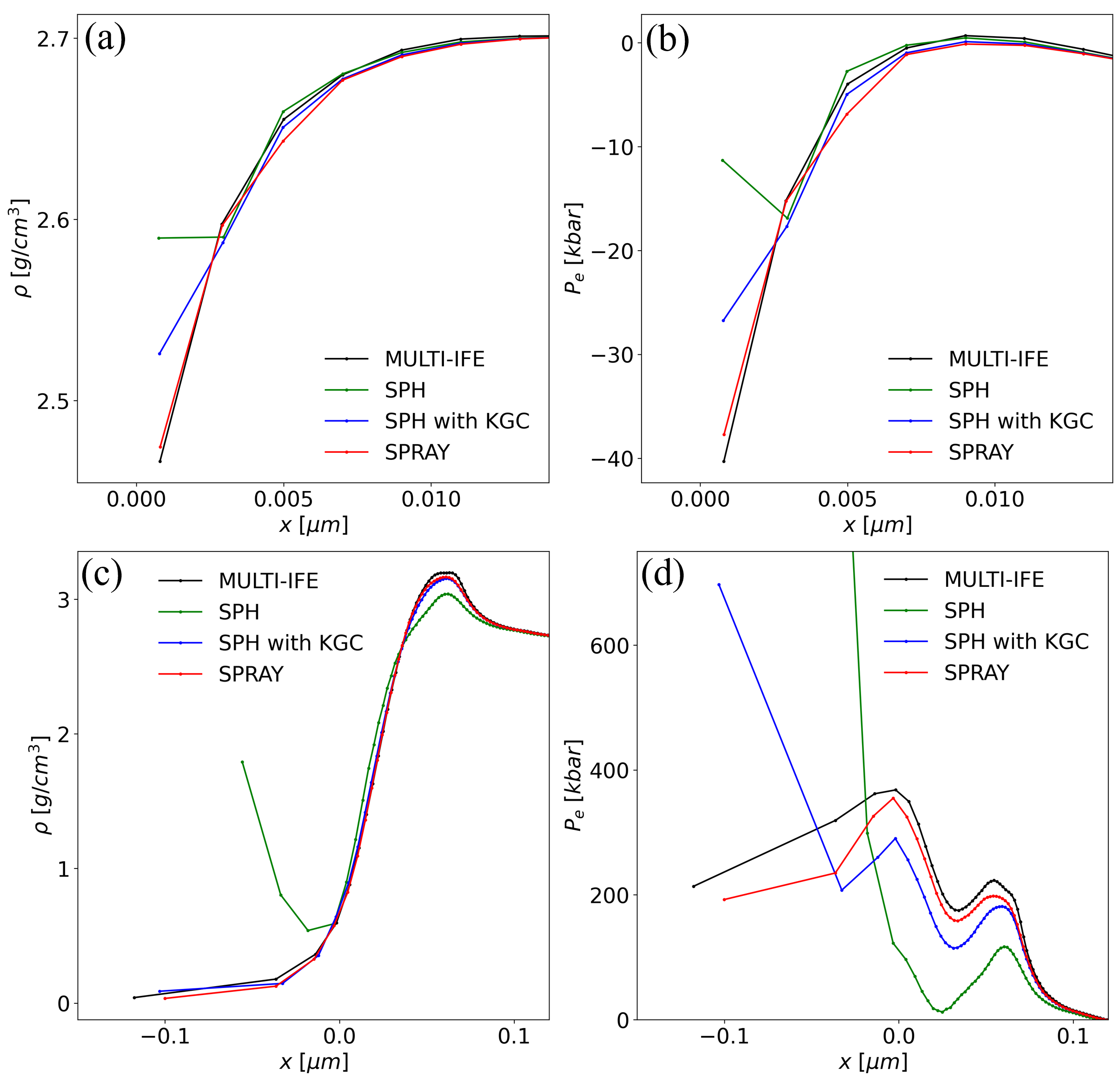}
    \caption{\textmd{Comparison of the free surface boundary treatment schemes. (a) and (b) are density and electron pressure profiles when $I_{peak}\sim10^{14}\ W/cm^2$ laser is incident on a solid metal target from the left side. (c) and (d) are the same profiles at later time step. Density is calculated by the continuity equation, and the electron pressure is determined from the equation-of-state data calculated by MPQEOS\cite{Kemp1998}. \hl{The negative electron pressure shown in (b) is due to the semiempirical chemical bonding correction of the QEOS model \mbox{\cite{More1988}}.} The reference profile is calculated by MULTI-IFE\cite{Ramis2016}\hl{, shown in black solid curve}.}}
    \label{fig:3}
\end{figure}

By applying this scheme, \hl{the accuracy is greatly improved near the vicinity of boundaries. Its effectiveness in tracking the density and pressure profiles is illustrated with a verification test (see \mbox{\textbf{Figure \ref{fig:3}}}). } Here, the irradiation of a solid metal target by a laser pulse with peak intensity $I_{peak}\sim10^{14} W/cm^2$ and pulse length $\tau_L\sim100 ps$ is simulated, where density is calculated with equation \eqref{2.1.3-36}. From the figure, it is immediately evident that the original SPH scheme and the SPH scheme with KGC \hl{(kernel gradient correction)} filter \hl{\mbox{\cite{Shao2012}}} applied cannot reliably reproduce the sharp falloff of density at the edge. This leads to a serious error in the simulation, since the electron pressure, which is highly sensitive to the plasma density, would be distorted, which would then yield inaccurate pressure gradient-driven velocity evolution, thus further distorting the density evaluation from the continuity equation. Hence, a robust and reliable scheme to handle freely expanding boundary surface is crucial in simulating high-intensity laser plasma problems.

\subsection{Laser energy deposition}\label{sec:2.5}

The absorption of long-pulse high-intensity laser in plasmas is dominated by the inverse bremsstrahlung absorption process \cite{Pfalzner2006}. The wave equation for a monochromatic, planar electromagnetic wave travelling in z-direction is as follows:

\begin{equation}\label{2.5-61}
    \frac{d^2E}{dz^2}+\frac{\omega^2\varepsilon}{c}E=0
\end{equation}
where $\omega$ is the laser angular frequency.

The dielectric function $\varepsilon$ is obtained by coupling the Maxwell's equations with the electron equation of motion:

\begin{gather}
    m\frac{d\boldsymbol{v}^e}{dt}=-e\boldsymbol{E}-m^e\nu^{ei}\boldsymbol{v}^e\label{2.5-62}\\
    \varepsilon=1-\frac{(\omega^p)^2}{\omega^2(1+i\nu^{ei}/\omega)}\label{2.5-63}
\end{gather}
where $\omega^p$ is the plasma frequency. Here, by assuming that the wavelength is shorter than the plasma gradient scale length, the Wentzel-Kramers-Brillouin (WKB) approximation can be applied \cite{Eliezer2003}. Although solving the full Maxwell equation\hl{s} could incorporate more detailed physics such as laser beam-beam interactions, the WKB approximation is commonly used as it is appropriate for the laser parameter range targeted in thus study (long pulse length in the order of nanoseconds) when modeling physical phenomena occurring on the hydrodynamic scale. Assuming a solution in the form

\begin{equation}\label{2.5-64}
    E(z)=E_0(z)\exp{\left(\frac{i\omega}{c}\int^z{\psi(\zeta)d\zeta}\right)}
\end{equation}

equation \eqref{2.5-61} yields
\begin{equation}\label{2.5-65}
    \frac{d^2E_0}{dz^2}+\left[\frac{i\omega}{c}\left(2\psi\frac{dE_0}{dz}+E_0\frac{d\psi}{dz}\right)+\frac{\omega^2E_0}{c^2}(\varepsilon-\psi^2)\right]=0
\end{equation}

The assumption that the zeroth-order terms are much greater than the higher-order terms yields the dispersion relation

\begin{equation}\label{2.5-66}
    k^2(z)=\frac{\omega^2\varepsilon}{c^2}=\frac{\omega^2}{c^2}\left(1-\frac{(\omega^p)^2}{\omega^2(1+i\nu^{ei}/\omega)}\right)
\end{equation}
and by Taylor expansion, the absorption coefficient is derived:
\begin{equation}\label{2.5-67}
    \kappa^{ib}=\frac{\nu^{ei}}{c}\frac{(\omega^p)^2}{\omega^2}\left(1-\frac{(\omega^p)^2}{\omega^2}\right)^{-1/2}
\end{equation}

With the absorption coefficient, the governing equation for the laser energy attenuation can be expressed as below:

\begin{equation}\label{2.5-68}
    \frac{d\phi}{dr}=-\kappa^{ib}\phi=-\frac{\nu^{ei}}{c}\frac{(\omega^p)^2}{\omega^2}\frac{\phi}{\sqrt{1-(\omega^p)^2/\omega^2}}
\end{equation}

\begin{figure}[htbp]
    \centering
    \includegraphics[width=0.6\linewidth]{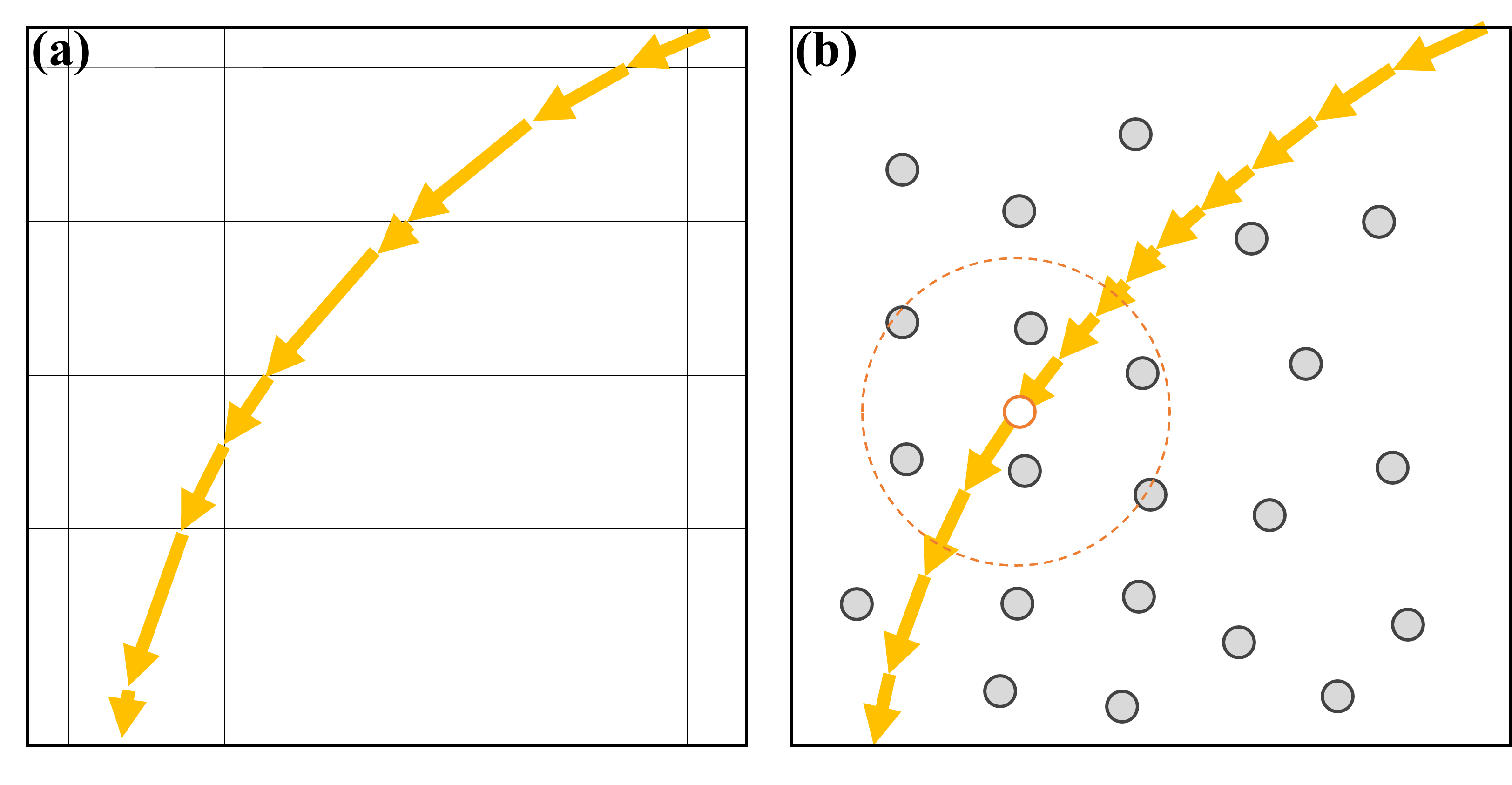}
    \caption{\textmd{Comparison between grid-based and mesh-free ray-tracing schemes. (a) illustrates the conventional grid-based ray-tracing method, whereas (b) depicts the mesh-free method implemented in SPRAY. In the mesh-free method, each ray is treated as a hypothetical SPH particle (hollow orange circle) traversing the electron density gradient field.}}
    \label{fig:6}
\end{figure}

In order to determine the amount of energy absorbed by individual particles, the laser ray trajectory needs to be calculated. Conventionally, grid-based ray-tracing schemes \mbox{\cite{Kaiser2000}} are used in state-of-the-art Eulerian codes \mbox{\cite{Fryxell2000}}. Such method\hl{s} assume \hl{a} parabolic trajectory within each cell, and an instantenous refraction on cell interfaces. Then, the track length within each cell is used to compute the absorbed energy. However, in order to preserve the mesh-free nature of SPH, a new ray-tracing scheme that does not rely on any underlying grid is developed (see \mbox{\textbf{Figure \ref{fig:6}}}).

In this scheme, each laser ray is treated as a hypothetical ray particle. Its equation of motion is given as the following:

\begin{equation}\label{2.5-69}
    \frac{dv}{dt} = -\frac{c^2}{2}\nabla\left(\frac{n^e}{n^c}\right)
\end{equation}
where $n^c$ is the critical density at which the laser wave is cut-off by the plasma wave ($\omega=\omega^p$). In order to determine the ray trajectory, the electron density gradient needs to be known. Hence, the electron density gradient is estimated for each fluid particle $a$ $\langle\nabla n_a^e\rangle=\sum_b{(m_b/\rho_b)n_b^e\nabla W}$, and it is SPH interpolated at the ray particle's position $\langle\nabla n_{ray}^e\rangle=\sum_a{(m_a/\rho_a)\nabla n_a^e W}$. Then, the ray advances accordingly by evaluating equation \mbox{\eqref{2.5-69}} for a fixed small $\Delta t$. The benefit of this approach is that, since ray velocity decreases where refraction occurs the most, the trajectory resolution is automatically refined near the reflection point (illustrated with differing arrow lengths in \mbox{\textbf{Figure \ref{fig:6}(b)}}).

The laser energy attenuation is computed based on the individual discrete track lengths obtained during the ray particle traversal. At each location, the absorption coefficient of nearby particles are gathered with standard SPH interpolation, and the deposited energy is then calculated with equation \mbox{\eqref{2.5-68}}. The energy is distributed to nearby particles based on the kernel weight of each particle. The energy absorbed by each particle corresponds to the last term in equation \mbox{\eqref{2.1.2-13}}.

\hl{The ray-tracing method described above differs from previous efforts to implement ray-tracing in SPH codes \mbox{\cite{Altay2008}}. In \mbox{\cite{Altay2008}}, a Monte Carlo ray-tracing algorithm for modeling radiative transfer is developed to sample photon packets along one dimensional characteristics with a predefined length. Although their proposed method is similar to the scheme described in this section in a sense that both approaches utilize SPH kernel interpolations, their ray-tracer does not consider refraction of the rays. This assumption is valid for modeling ionizing photons, but is not justifiable for modeling lasers propagating in a medium with steep electron density gradients.}

The ray tracing routine is \hl{executed} on CPUs instead of GPUs, so that it \hl{can} be executed concurrently with the SPH calculations on GPUs. \hl{This is particularly beneficial when the computational load of the ray tracing routine is high due to large number of rays or irregular electron density profiles.} It is parallelized using POSIX threads, and the number of threads can be specified as an input. \hl{Because laser beam-beam interaction effects such as cross-beam energy transfer (CBET) are not considered, the rays can be traced independently from each other, which allows for highly parallelized implementation.} The laser is specified as a set of beams, and each beam consists of multiple rays. If the beam width and the number of rays are determined, each ray is designated a "launch zone." This zone is obtained by dividing the beam width into equally spaced zones for each ray. The actual ray launch position is randomly selected within its launch zone. This technique is implemented to compensate for the unphysical effects due to \hl{the} finite number of rays used. If the ray launch position is fixed, it is likely that certain particles would be preferentially heated due to them being close to the ray trajectory. By uniformly sampling the ray launch position, all particles within the width of the beam would be heated without discrimination.

\subsection{Nearest Neighbor Particle Search (NNPS)}\label{sec:2.6}

In SPH algorithm, a majority of computational time is spent on searching and identifying neighboring particles for each particle, a process called nearest neighbor particle search (NNPS). As the number of particles ($N$) increases, if the relative distances between the particles are calculated as is, the complexity is $O\left(N^2\right)$. This poses a serious issue since the computational load increases dramatically as greater number of particles are used, and considering that this search process needs to be performed at every time step, the computational burden quickly becomes unrealistic. A number of studies have suggested various numerical algorithms to efficiently search for neighboring particles, including Barnes-Hut algorithm-based methods such as the hierarchical tree method and the kd-tree method \hl{\mbox{\cite{Hernquist1989,Barnes1986,Hernquist1987,Price2018,Gafton2011}}}. However, these approaches share a major drawback, which is that \hl{the tree (re)construction needs to be performed at every time step, but implementing it in a manner suitable for GPU parallelization is not trivial.}

SPRAY adopts a cell-based approach, which is one of the popular choices of modern SPH codes \cite{Xia2016}. The simulation domain is subdivided into hypothetical cells, and the index of the cell that each particle resides in can be easily computed. The dimensions of these cells are automatically determined based on the average smoothing length of the SPH particles during the initialization phase. Then, the particles are sorted based on their cell indices, so that the particles in any arbitrary cell can be identified directly.  During the neighboring particle search phase in the SPH calculation, only the particles that reside in nearby cells are searched, which dramatically reduces the computational load. The search range of nearby cells is determined based on the ratio of the particle's smoothing length to the cell dimensions, which guarantees that all neighboring particles are considered. Just like the tree-based methods, the complexity is maintained as $O\left(Nlog{N}\right)$, but it is still fully parallelizable. Furthermore, because the smoothing length is tied to the particle density, passive load balancing is achieved. Consider a cell with a large number of particles inside. This would suggest that the density of those particles would be high, which implies that the smoothing length is small. Then, when searching for neighbors in nearby cells, only a few cells will be searched. In contrast, for a cell with small number of particles, the particles would have low density and large smoothing length, which in turn results in searching more cells that are farther away. Therefore, the computational load is roughly similar across different particles. Nevertheless, the overall efficiency and performance would be quite sensitive to the general particle distribution compared to aforementioned tree-based approaches.

\begin{figure}[htbp]
    \centering
    \includegraphics[width=0.4\linewidth]{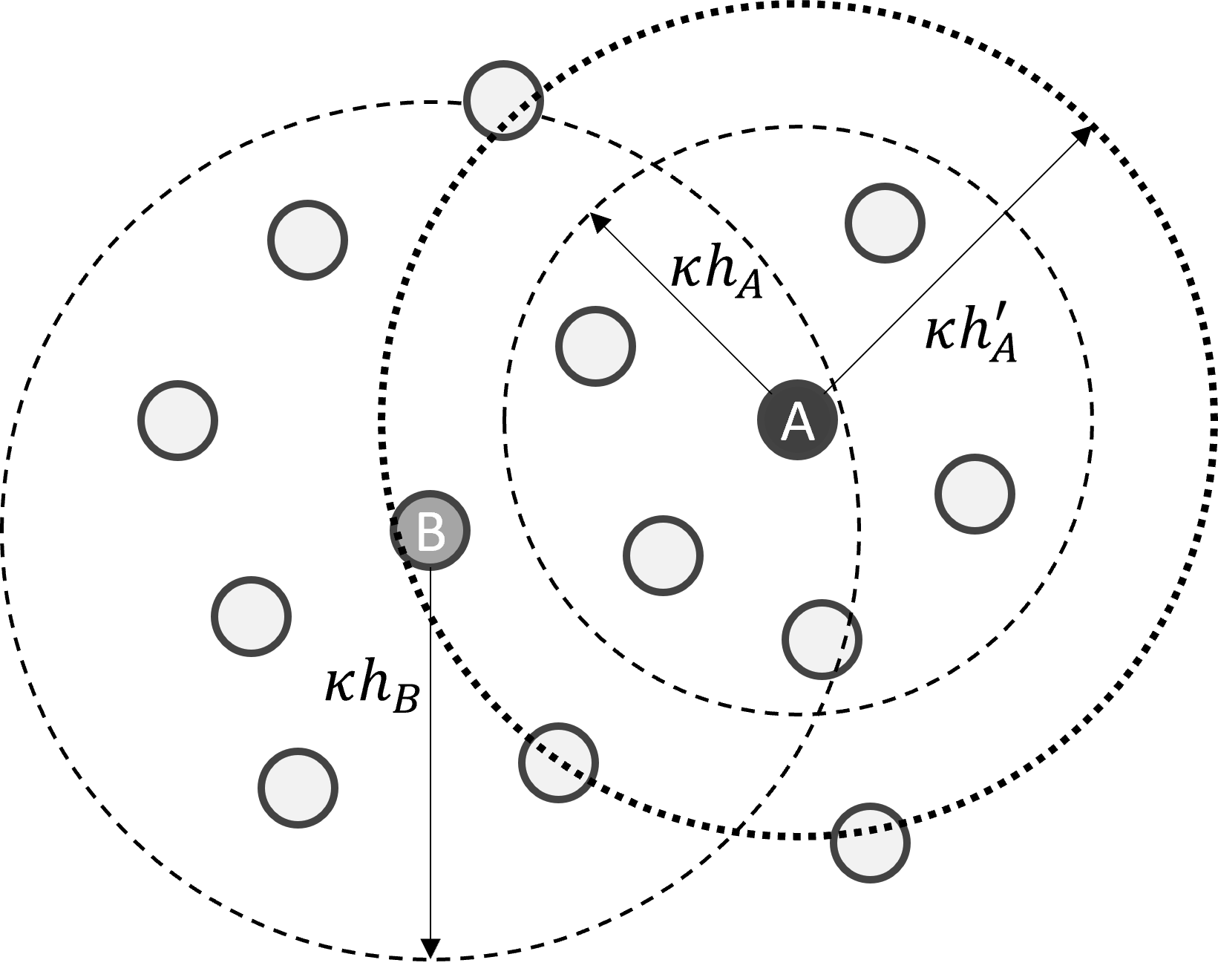}
    \caption{\textmd{Illustration of the search range optimization scheme. Due to the density differences, particle A does not "see" particle B within its smoothing range $\kappa h_A$, but particle B does include particle A in its smoothing length $\kappa h_B$. Therefore, the nearby particles of A are scanned to define a smoothing range $\kappa h_A^{\prime}$ specifically for the purpose of complete search of neighbors.}}
    \label{fig:8}
\end{figure}

Due to the concurrent occurence of ablation and compression, the density of the particles could vary by several orders of magnitude in laser-plasma interaction simulation. If the initial masses of the particles are uniform, this entails large discrepancies in particle volumes and thus the smoothing lengths. In this case, the coexistence of particles with drastically different smoothing lengths could pose an issue where one particle sees another particle within its smoothing length $h$, but not the other way around (see \textbf{Figure \mbox{\ref{fig:8}}}). In order to guarantee interaction symmetry, one has to \hl{devise a} method to ensure all neighbors are mutual. The easiest yet inefficient solution is to determine the maximum smoothing length $h_{max}$ at each time step, and use that value to search for neighbors whenever SPH calculation is needed. In order to avoid this, a separate smoothing length $h^{\prime}$ dedicated for searching neighbors is defined, while the original smoothing length $h$ remains tied to the particle volume, serving as a variable knob for resolution control. This second smoothing length $h^{\prime}$ is computed by first identifying the farthest particle that includes itself as a neighbor, then by multiplying a constant slightly greater than unity to that distance (and divided by the factor $\kappa=2$). This method of search range optimization by tracking two smoothing lengths (one for neighbor search, another for numerical resolution) enhances the computational efficiency for simulations with dramatic differences in particle densities and volumes.

\begin{figure}[htbp]
    \centering
    \includegraphics[width=0.6\linewidth]{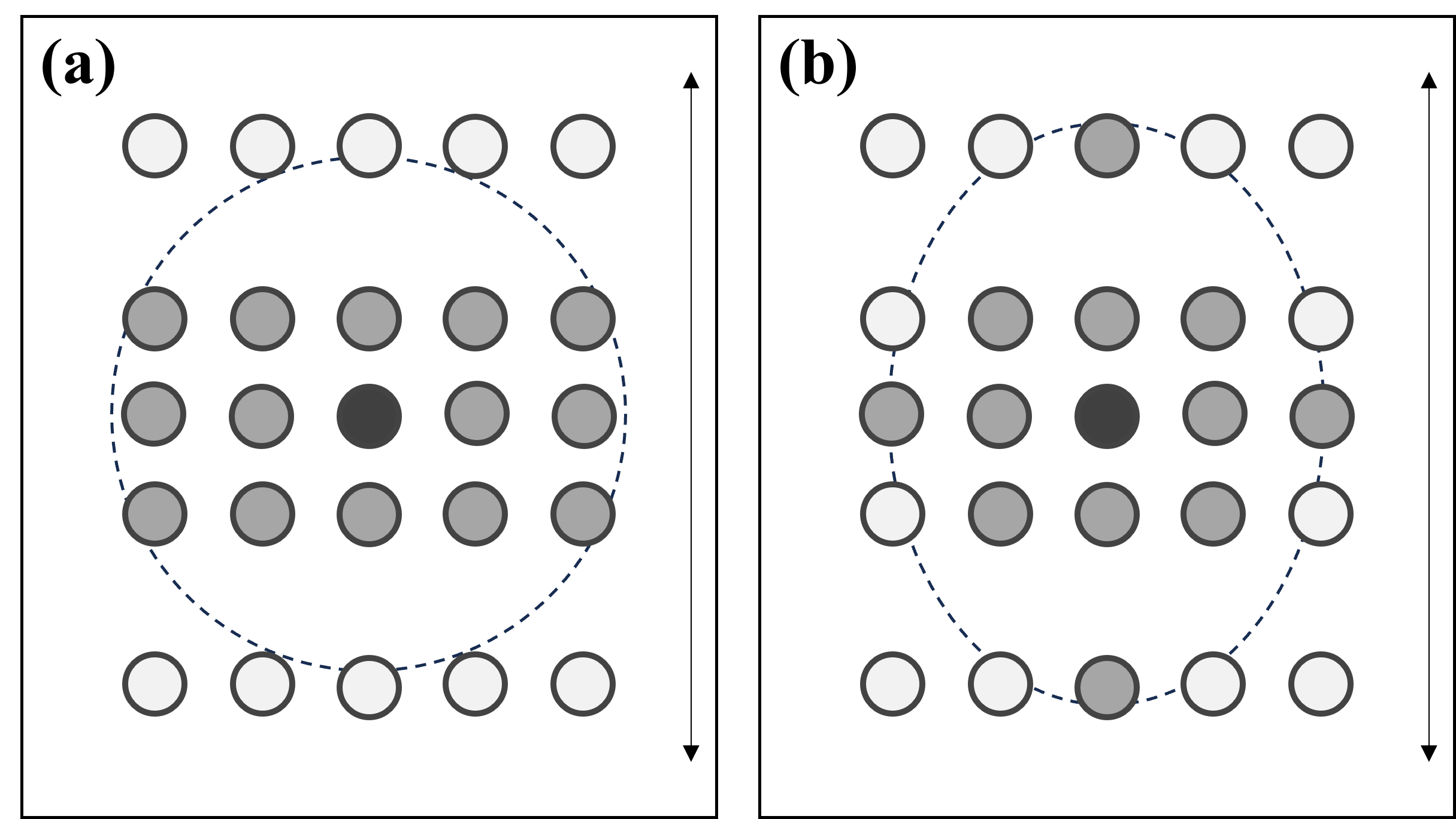}
    \caption{\textmd{Depiction of anisotropic kernel. When a rapid directional expansion of fluid such as laser ablation occurs, the particle distribution would be modified to resemble the situation illustrated in this figure. If the usual isotropic kernel is used (case (a)), the neighbors in the direction of expansion (indicated by arrow) contribute less to the SPH interpolation compared to the particles in the opposite direction. By using anisotropic kernel (case (b)), a more balanced group of neighbor particles are selected.}}
    \label{fig:9}
\end{figure}

In addition, in the presence of laser-induced ablation, a strong expansion in the direction of laser incidence occurs. As a result, a noticeable anisotropy in particle distribution is formed where the interparticle separation is much greater in the direction of ablative expansion compared to the perpendicular direction (see \textbf{Figure \mbox{\ref{fig:9}}}). This prompted the use of \hl{an} anisotropic kernel to ensure balanced particle interaction in terms of directionality. The ellipsoidal kernel proposed by \mbox{\cite{Owen1998}} is implemented, and the shape and size of the kernel are evolved based on the gradient of the velocity field (defined as the deformation tensor), thus naturally adapting to the directionality of expansion and compression. Because the general direction of ablative expansion can be anticipated, the initial tilting angle of the ellipsoid kernel is set to the surface normal direction during the particle initialization phase. Furthermore, the kernel gradient correction (KGC) filter is applied to the calculation of the deformation tensor to improve the accuracy and reliability of this method\mbox{\cite{Shao2012}}.

\subsection{GPU parallelization}\label{sec:2.7}

\begin{figure}[htbp]
    \centering
    \includegraphics[width=0.4\linewidth]{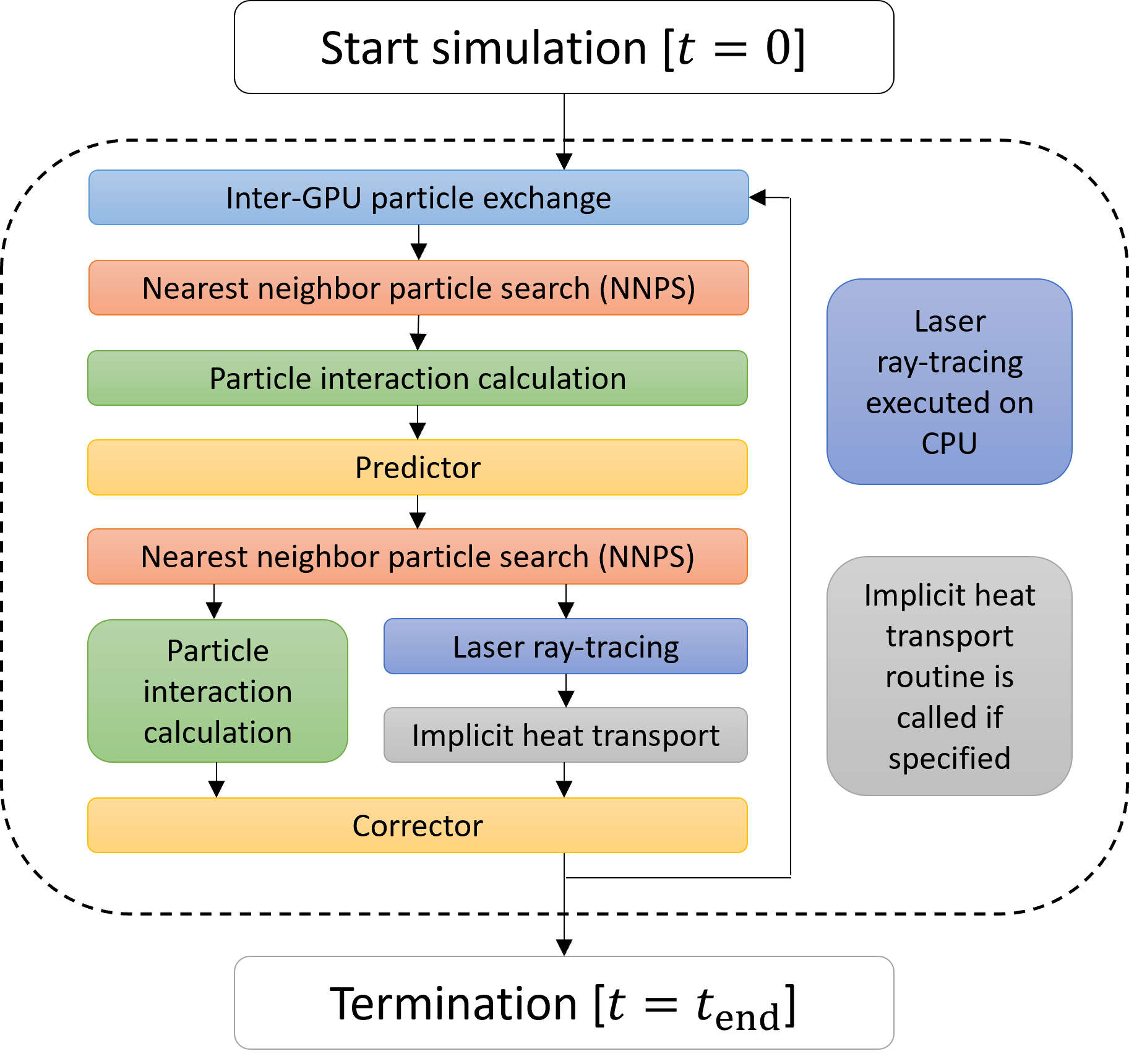}
    \caption{\textmd{SPRAY code workflow. The steps within the dotted box are repeated for each time step. When multi-GPU capability is enabled, inter-GPU particle exchange is executed at the beginning of each step. Implicit heat transport is also enabled if requested in the input file. The laser ray-tracing routine is executed on the CPU side so that it could run concurrently with the SPH calculations on the GPUs.}}
    \label{fig:10}
\end{figure}

SPRAY is a CUDA-based GPU parallelized code that can utilize multiple NVIDIA GPUs for its SPH calculations. In order to maximize the simulation performance, meticulous design of algorithms and optimizations is imperative. In general, each CUDA thread carries out relevant SPH calculation for one particle, looping over neighbor particles by utilizing the nearest neighbor particle search (NNPS) scheme described in the previous section. The overall workflow of each time step is illustrated in \mbox{\textbf{Figure \ref{fig:10}}}.

\begin{figure}[htbp]
    \centering
    \includegraphics[width=0.4\linewidth]{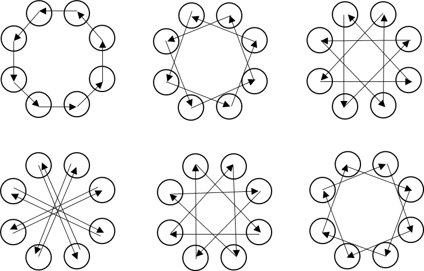}
    \caption{\textmd{Illustration of GPU memory exchange order. To maximize memory throughput, the order of memory operations is optimized so that there is no overlap in utilization of memory transfer channels. The case illustrated in this figure assumes eight GPUs, and at each stage of memory operation, each GPU both sends and receives data concurrently. The grouping method is shown. The last stage, which is the exact reverse of the first stage, is not included.}}
    \label{fig:11}
\end{figure}

In SPRAY, a predictor-corrector based time integration is used, and the workflow for each time step is organized accordingly. If multiple GPUs are being utilized, \hl{an} inter-GPU particle exchange routine is called at the start of each time step to synchronize all particle information across the GPUs. Because no assumption of the geometry of the target problem is made, although each GPU is responsible for solving the governing equations for a certain portion of the particles, it needs to gather all particles' information. Hence, if there are $N$ particles and $M$ GPUs, each GPU needs to send and receive $N/M$ particles' information $M-1$ times. Because the GPU memory bandwidth is the most important limiting factor, a specialized algorithm is developed to optimize the order in which the GPUs exchange data (see \mbox{\textbf{Figure \ref{fig:11}}}). In short, the GPUs execute memory copy operations in a specific order while ensuring that no memory transfer is overlapped on the same channel.

\begin{figure}[htbp]
    \centering
    \includegraphics[width=0.95\linewidth]{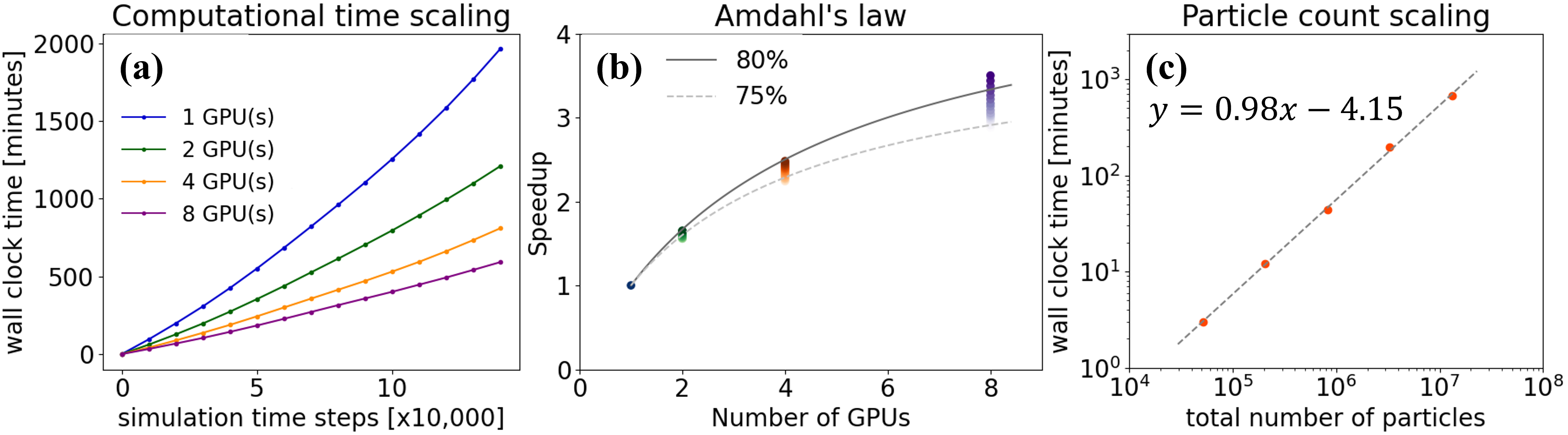}
    \caption{\textmd{Scaling study results. (a) is the elapsed computation time as the simulation progresses. The reduction in the wall clock time as the number of GPU increases is clearly demonstrated. (b) is comparison with the Amdahl's law curve. From the fitted curve, the portion of the code that benefits from parallelization is greater than 75\% (indicated by the light gray dashed curve). The portion occupied by the parallelized tasks seems to increase as simulation progresses (indicated by the darker shade in color). (c) is the particle number scaling test result. As the number of particles increases to over 10 million particles, the computation time scales well. The linear fit of the log-log plot is close to unity.}}
    \label{fig:12}
\end{figure}

The performance scaling of the code is tested. The test case used for this benchmark is the implosion simulation described in \mbox{\textbf{Section \ref{sec:3.4}}}. First, the computation time is measured as the number of GPUs increased (\mbox{\textbf{Figure \ref{fig:12}(a)}}). The reduction in computation time as more GPUs are used is clear, and it is worth noting that the speedup increases as the simulation progresses. This is because at latter stage\hl{s} of simulation, more numerical schemes start to take effect to resolve the physical phenomena (e.g. numerical dissipation switch activation), so the percentage of the time taken up by memory operation overhead starts to decrease. This trend is more evident in the Amdahl's law test (\mbox{\textbf{Figure \ref{fig:12}(b)}}). Amdahl's law predicts the theoretical speedup of a program depending on the proportion of the code that benefits from parallelization. Judging from the results, at least 75\% of the code is accelerated by the parallelization scheme. As discussed, the speedup increases as simulation progresses (data points with darker shade correspond to later stages of simulation), and the parallelized portion seems to exceed 80\%.

The computation time at various particle numbers is also tested (\mbox{\textbf{Figure \ref{fig:12}(c)}}). The number of particles is increased from the order of 10 thousand to 10 million. The linear fit of the data points in the log-log plot is close to unity, which suggests close to $O(n)$ scaling. If the total particle count \hl{increases} indefinitely, it would eventually converge to $O(n \log n)$ complexity due to the NNPS scheme. From these results, it is evident that the GPU parallelization of SPRAY is exhibiting expected scaling behavior.

\section{Code verification}

\subsection{Sod shock tube problem - Hydrodynamics with shock wave capturing capability}\label{3.1}

One of the main challenges numerical codes that simulate high-intensity laser dynamics need to address is the presence of intense shock waves. When a target is irradiated with a high power laser, a series of shock waves are created and propagate into the target, resulting in a high compression domain located behind the laser-plasma interaction region. Since shock waves travel faster in the lower density medium, the wave fronts eventually pile up at a localized point, which then induce sharp gradients in both density and velocity profiles. Due to the gradient scale length typically being much smaller than the numerical spatial resolution, the presence of intense shock waves could cause unphysical oscillations. Therefore, as described in \textbf{Section \ref{sec:2.3}}, artificial viscosity is implemented to address this issue. Here, the robustness of the scheme is verified with a well-known hydrodynamics benchmark problem called Sod shock tube test \cite{Sod1978}.

\begin{figure}[htbp]
\begin{subfigure}{\linewidth}
    \centering
    \includegraphics[width=0.8\linewidth]{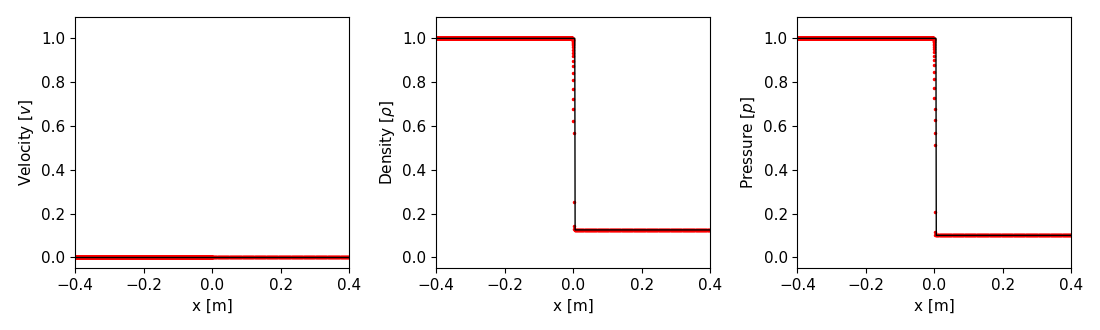}
    \caption{Initial configuration}
    \label{fig:subfig1}
\end{subfigure}
\begin{subfigure}{\linewidth}
    \centering
    \includegraphics[width=0.8\linewidth]{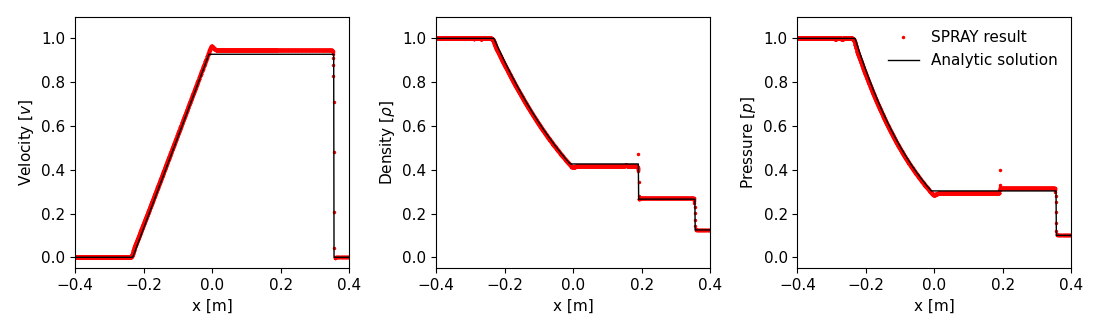}
    \caption{Calculation result}
    \label{fig:subfig2}
\end{subfigure}
    \caption{\textmd{Sod shock tube problem benchmark results. Top row illustrates the initial conditions for each variable. Bottom row compares the calculation results (red dots) with the analytic exact solution (black solid line). }}
    \label{fig:13}   
\end{figure}

Sod shock tube problem, named after G. A. Sod who introduced the problem, is a well-established benchmark problem for Riemann solvers. The initial setup is comprised of two regions on either side of the origin, and they differ in the initial density and pressure, with zero velocity (initial conditions are listed in \textbf{Table \ref{t1}}). The simulation results, along with the analytical solutions, for density, pressure and velocity are presented in \textbf{Figure \ref{fig:13}}.

\begin{table}[htbp]
\centering
\caption{\textmd{Initial conditions for Sod shock tube benchmark. All physical parameters are normalized, dimensionless values.}}
 \label{t1}
\begin{tabular}{c c c c c}
    \toprule
     & Density [$\rho$] & Pressure [$p$] & Internal energy [$u$] & Velocity [$v$] \\
    \midrule 
    Left & 1.0 & 1.0 & 2.5 & 0.0 \\
    Right & 0.125 & 0.1 & 2.0 & 0.0 \\
    \bottomrule
\end{tabular}
\end{table}

Equation of state for ideal gas ($p=(\gamma-1)u\rho$) is used ($u$ is the specific internal energy) with the adiabatic index $\gamma=1.4$ , and the profile discontinuities are smoothed out in the preprocessing phase. Initial particle spacings were also adjusted to ensure consistency with the density estimation (i.e. each particle volume was ensured to satisfy the following relation $V=m/\rho$) \cite{Monaghan1997}.

\subsection{Long pulse laser irradiation of aluminum target - HEDP laser energy deposition coupling capability}\label{sec:3.2}

\begin{figure}[htbp]
    \centering
    \includegraphics[width=0.8\linewidth]{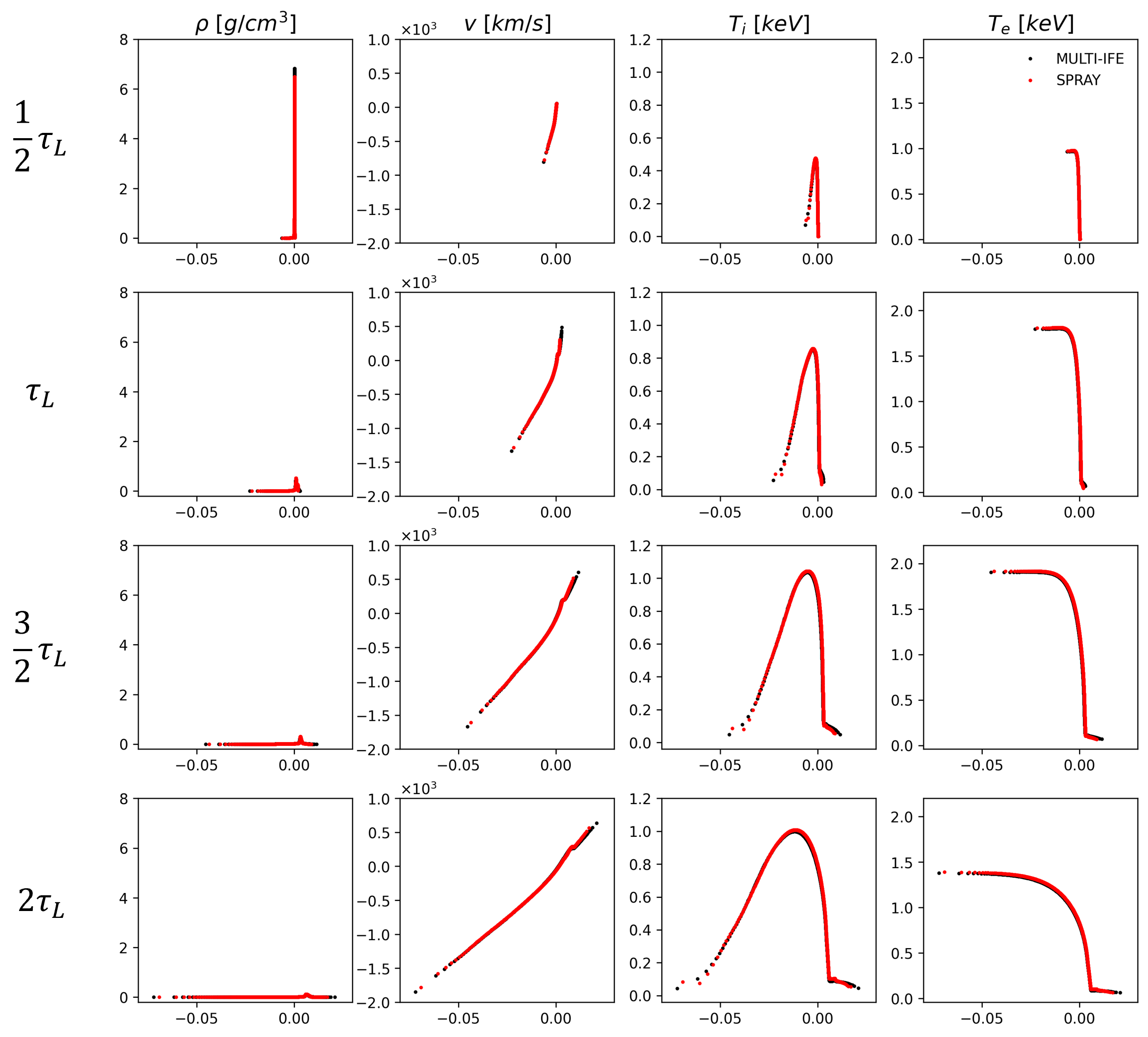}
    \caption{\textmd{Laser irradiation of aluminum target benchmark results at four time points ($\frac{1}{2}\tau_L,\tau_L,\frac{3}{2}\tau_L,2\tau_L$). SPRAY profiles are drawn in red dots, and MULTI-IFE\cite{Ramis2016} results are in black solid lines. The initial target position is [$0,2$], and the laser is incident from the left. The $x$-axis denotes the spatial coordinate in the unit of [$\mu m$]. The SPRAY and MULTI-IFE profiles are overlapped due to yielding almost identical results.}}
    \label{fig:14}   
\end{figure}

SPRAY is targeted towards simulations of laser-plasma interaction, and its accuracy and validity is verified by benchmarking the code with another code. Here, MULTI-IFE \cite{Ramis2016}, a cell-based Lagrangian code, is chosen to be the reference code. MULTI-IFE is an implicit radiation hydrodynamics code designed to study inertial fusion energy microcapsules, and it features laser ray-tracing and energy deposition, radiative and thermal energy transfer, as well as deuterium-tritium thermonuclear burning. For the benchmark, a cold aluminum target with $2\ \mu m$ of thickness is irradiated with a $3\times{10}^{14}\ W/cm^2$ laser pulse with wavelength ($\lambda_L$) of $440\ nm$, pulse duration (denoted as $\tau_L$) of $300\ ps$, and sine-squared pulse shape. The results are shown in \textbf{Figure \ref{fig:14}}. From the figure, it is quite difficult to tell apart the SPRAY and MULTI-IFE results due to the results from SPRAY being in close agreement with that from MULTI-IFE. This is a testament to the robustness of not only the laser energy deposition coupling scheme of the code, but also the radiation transport modeling implementation of SPRAY.

\begin{figure}[htbp]
\begin{subfigure}{\linewidth}
    \centering
    \includegraphics[width=0.6\linewidth]{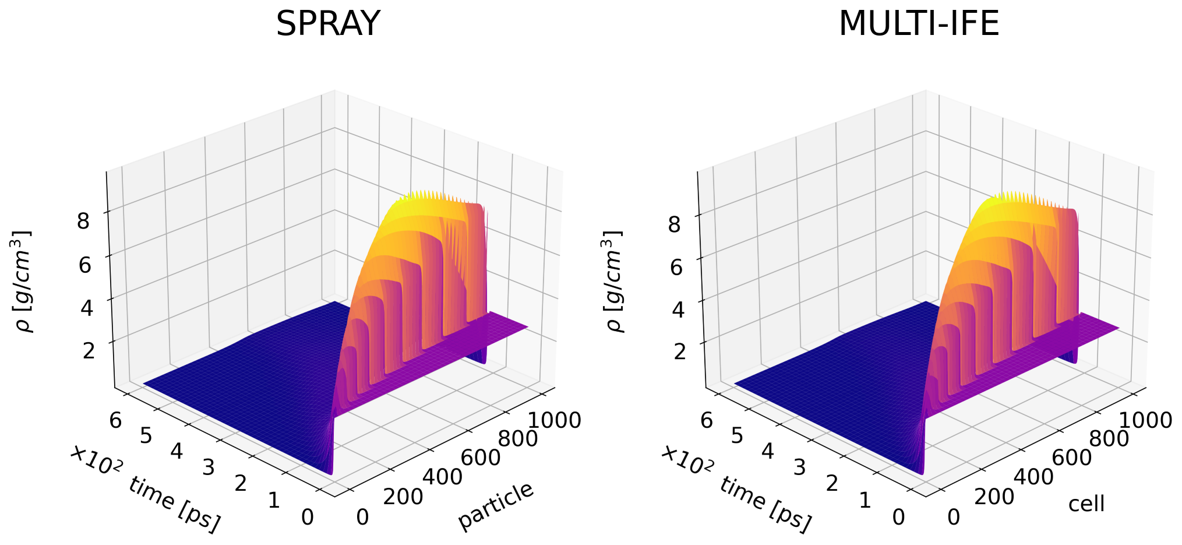}
    \includegraphics[width=0.6\linewidth]{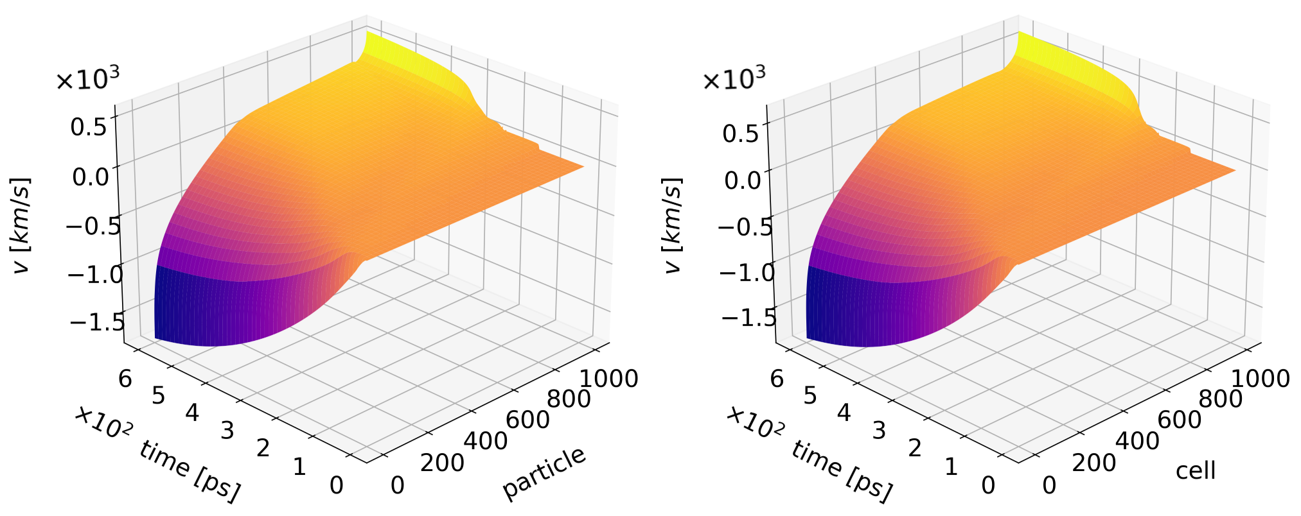}
\end{subfigure}
\begin{subfigure}{\linewidth}
    \centering
    \includegraphics[width=0.6\linewidth]{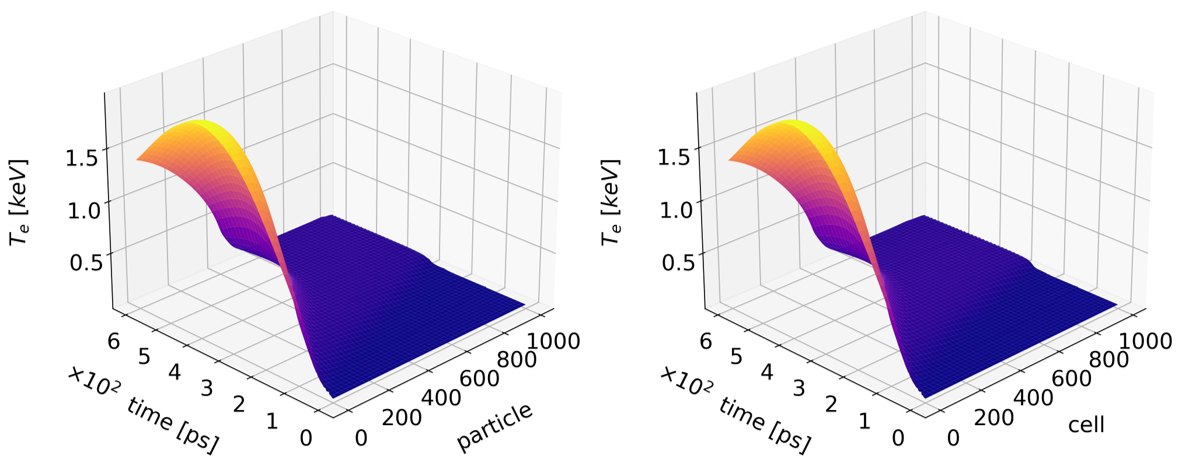}
    \includegraphics[width=0.6\linewidth]{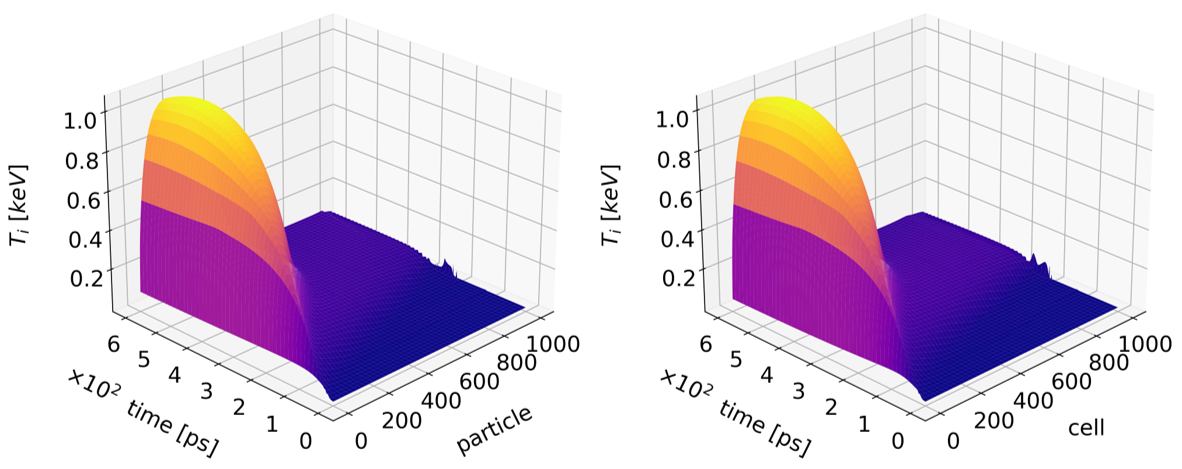}
\end{subfigure}
\begin{subfigure}{\linewidth}
    \centering
    \includegraphics[width=0.6\linewidth]{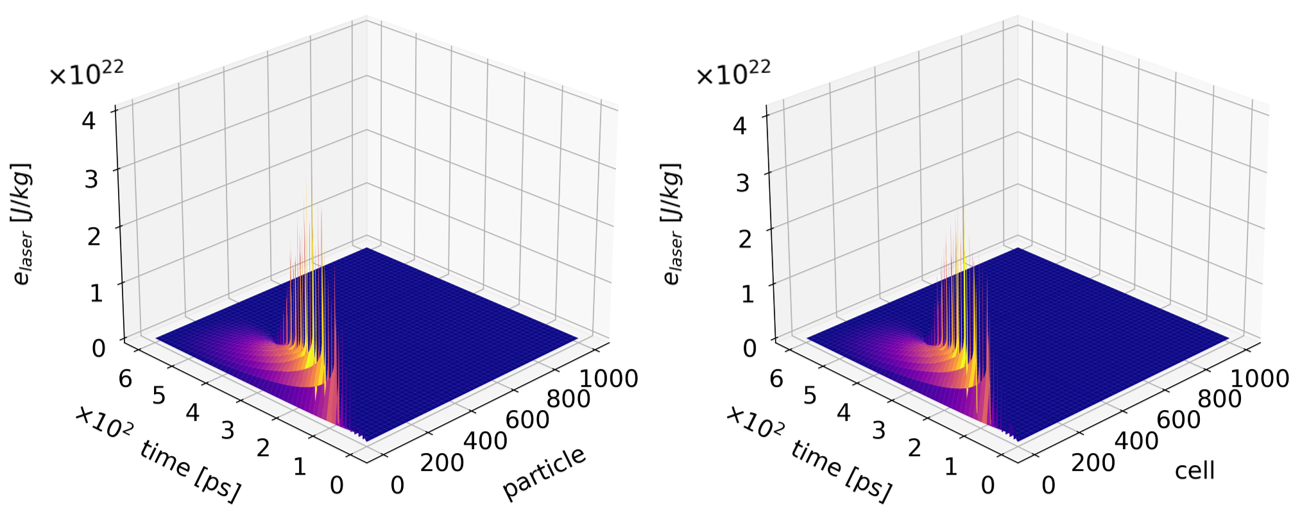}
\end{subfigure}
    \caption{\textmd{Comparison of SPRAY results with MULTI-IFE results. From the top, mass density, velocity, electron temperature, ion temperature and specific laser energy deposition of each particle/cell at each time are plotted. Because the same number of particles/cells are used in both codes, particle-to-cell correspondence is assumed.}}
    \label{fig:15}
\end{figure}

MULTI-IFE simulation configuration was composed of uniformly-sized 1,000 cells, while SPRAY used 1,000 SPH particles. Identical equation of state and opacity data were used in both codes (equation of state data from MPQEOS \cite{Kemp1998}, ionization and opacity data from SNOP \cite{Eidmann1994}). The electron-phonon collision frequency option was used in MULTI-IFE in order to match the SPRAY scheme. \hl{Heat transport equations were solved via the explicit time integration scheme with the maximum time step size capped at $0.1 fs$, and radiation transport was evaluated via the Jacobi-iteration scheme described in \mbox{\textbf{Section \ref{sec:2.4}}}.} \hl{The upper bound of the number of iterations required for the radiation transport solver to converge was in the order of a few hundred, peaking around 690 in this particular run.}

The overview of the profile evolutions is presented in \textbf{Figure \ref{fig:15}}. Density, velocity, ion/electron temperatures, and specific laser energy deposition rate for each cell/particle in MULTI-IFE/SPRAY are compared. By comparing the corresponding counterparts, it is evident that SPRAY code is capable of reproducing the results of MULTI-IFE code.

\begin{figure}[htbp]
    \centering
    \includegraphics[width=0.8\linewidth]{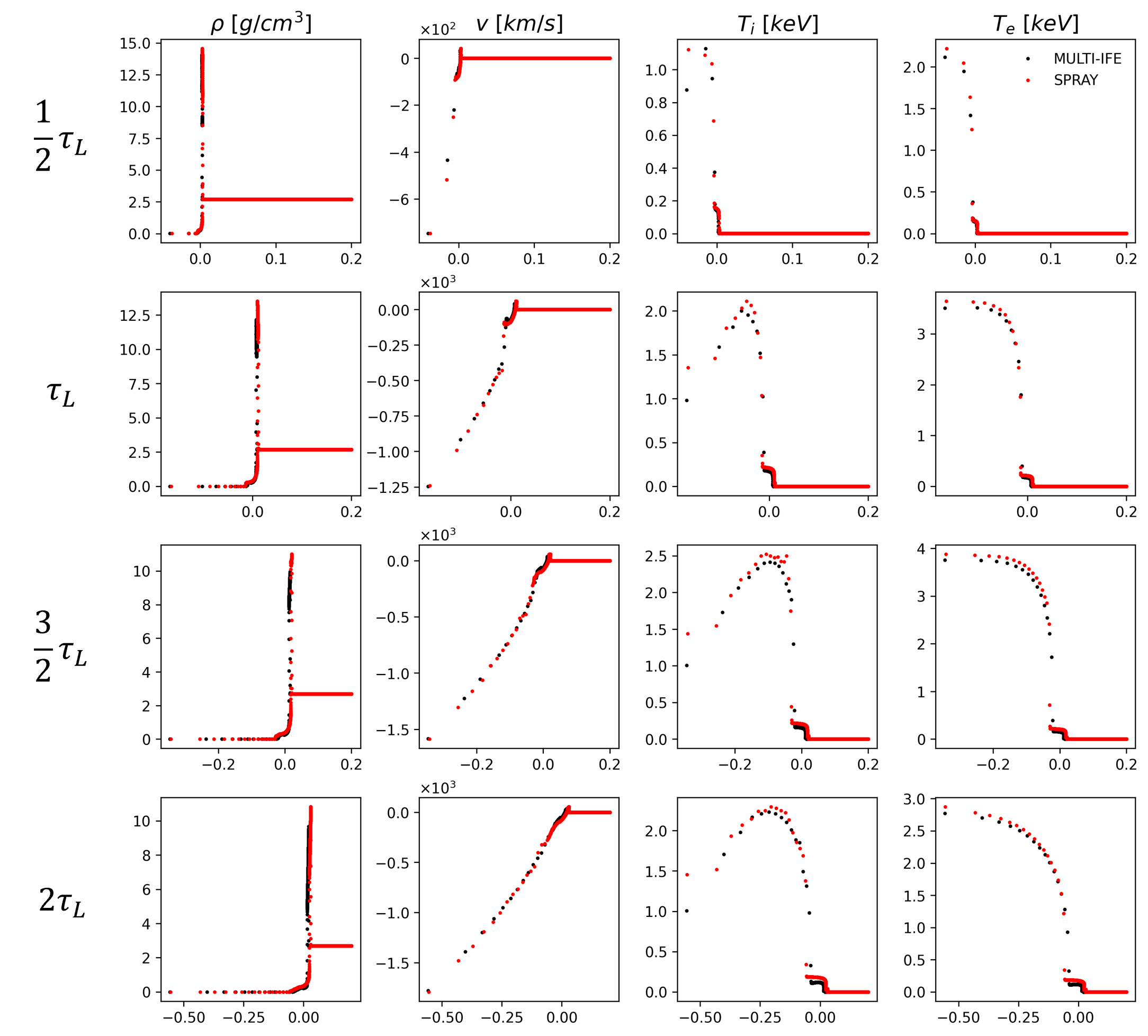}
    \caption{\textmd{Laser irradiation of an aluminum target benchmark results at four time points ($\frac{1}{2}\tau_L,\tau_L,\frac{3}{2}\tau_L,2\tau_L$). SPRAY profiles are drawn in red dots, and MULTI-IFE\cite{Ramis2016} results are in black solid lines. The initial target position is [$0,2\times10^{-1}$], and the laser is incident from the left. The $x$-axis denotes the spatial coordinate in the unit of [$cm$].}}
    \label{fig:16}  
\end{figure}

Another simulation with a different laser specification is also carried out. Instead of $\tau_L=300ps$ laser, an ICF relevant laser with the pulse length of $5ns$ and peak intensity of $5\times 10^{14} W/cm^2$ is simulated ($\lambda_L=350nm$). The target thickness is increased to $2mm$. The results are presented in \textbf{Figure \ref{fig:16}} in an identical manner.

\subsection{Rayleigh-Taylor instablity - Multidimensional hydrodynamic instability analysis capability}\label{sec:3.3}

\begin{figure}[htbp]
    \centering
    \includegraphics[width=0.8\linewidth]{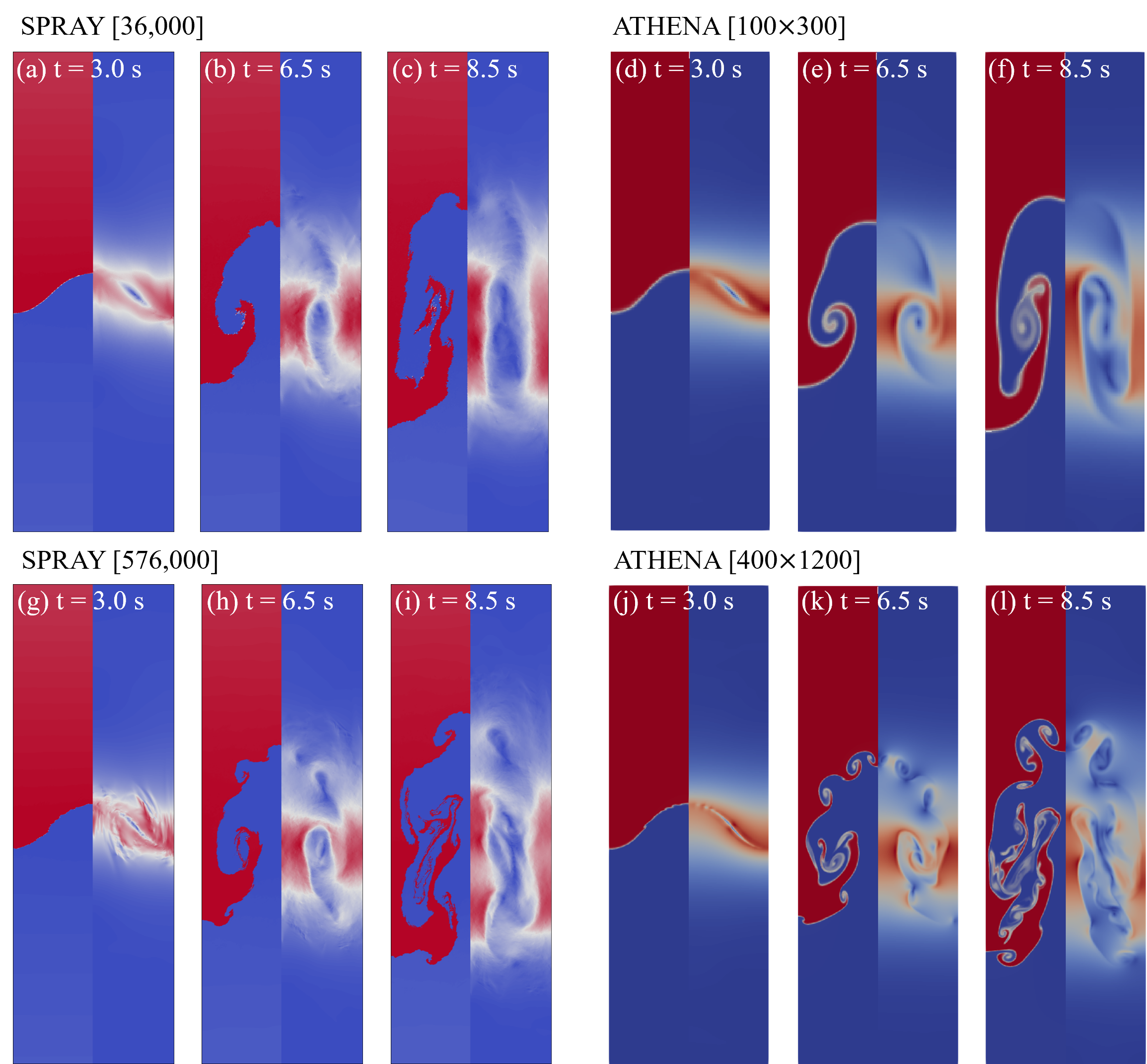}
    \caption{\textmd{\hl{Rayleigh-Taylor instability benchmark results. Top row corresponds to simulation results with 100x300 base resolution for ATHENA, and 36,000 particles for SPRAY. (a)-(c) are results of SPRAY at $t=3.0\ s$, $6.5\ s$, and $8.5\ s$, and (d)-(f) are ATHENA results at the same time points. The left half of each figure is density, and the right half is velocity.} \hl{Bottom row corresponds to 400x1200 base resolution for ATHENA, and 576,000 particles for SPRAY. (g)-(i) are results of SPRAY at $t=3.0\ s$, $6.5\ s$, and $8.5\ s$, and (j)-(l) are ATHENA results at the same time points.}}}
    \label{fig:17}
\end{figure}

\begin{figure}[htbp]
    \centering
    \includegraphics[width=0.9\linewidth]{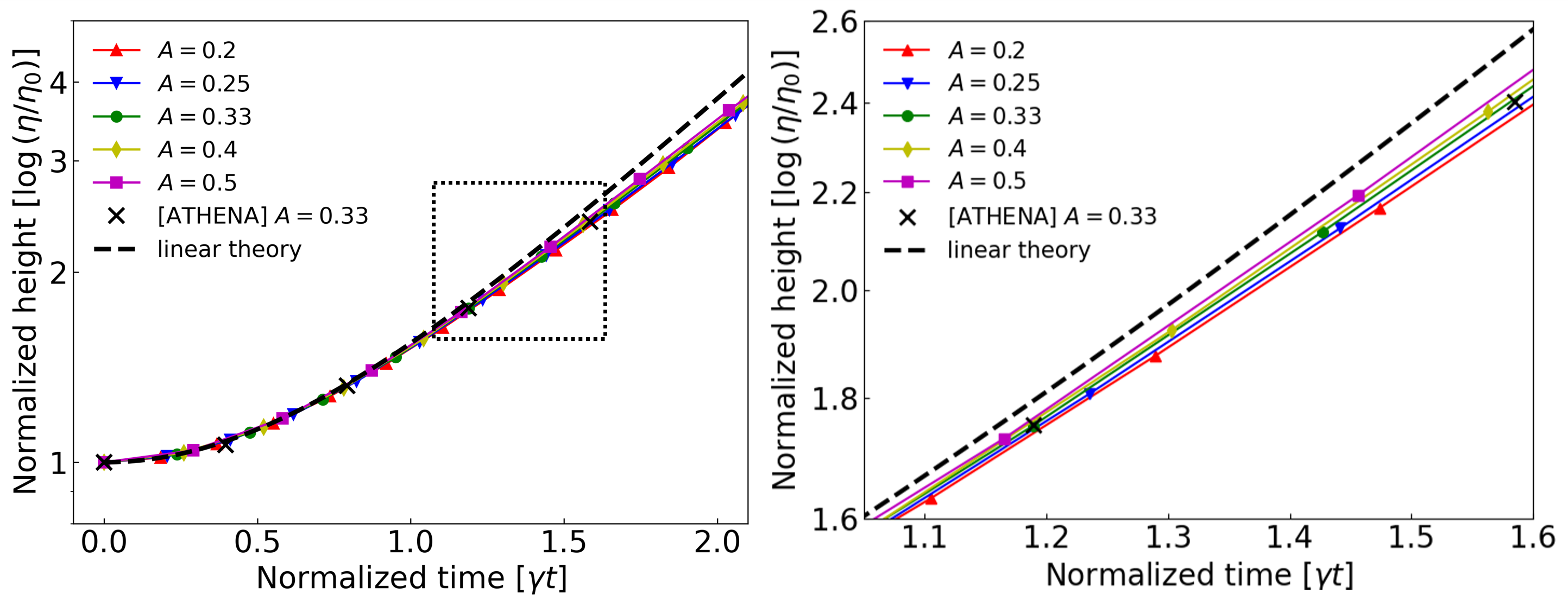}
    \caption{\textmd{Rayleigh-Taylor instability growth rate benchmark results. Each colored line illustrates SPRAY results with differing Atwood number $A$. Red line corresponds to $A=0.2$, blue line is for $A=0.25$, green line is for $A=0.33$, yellow line is for $A=0.4$, and purple line is for $A=0.5$. The dashed line is the analytic linear theory \cite{Mikaelian2014}, and the markers are ATHENA data points for $A=0.33$. The figure on the right is the zoomed-in version of the dashed-box region marked in the figure on the left.}}
    \label{fig:21}
\end{figure}

\begin{figure}[htbp]
    \centering
    \includegraphics[width=0.9\linewidth]{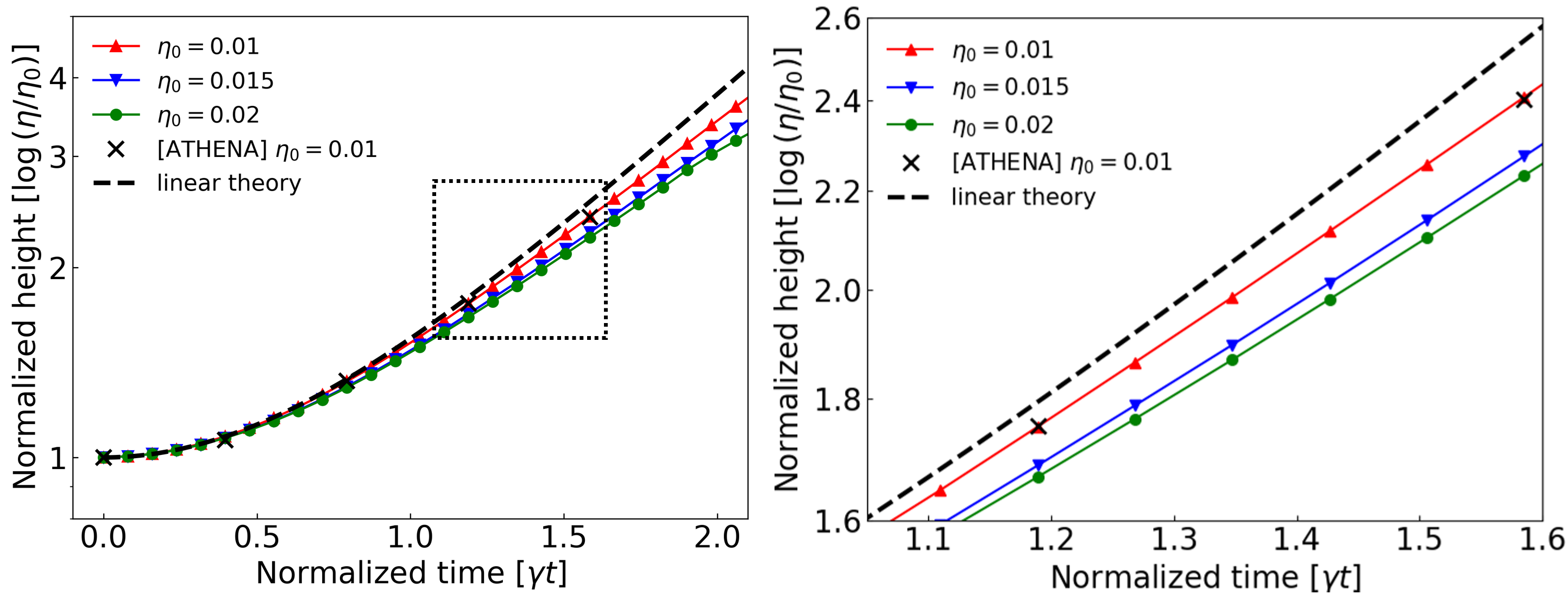}
    \caption{\textmd{Rayleigh-Taylor instability growth rate benchmark results. Each colored line illustrates SPRAY results with differing initial perturbation magnitude $\eta_0$. Red line corresponds to $\eta_0=0.01$, blue line is for $\eta_0=0.015$, and green line is for $\eta_0=0.02$. The dashed line is the analytic linear theory \cite{Mikaelian2014}, and the markers are ATHENA data points for $\eta_0=0.01$. The figure on the right is the zoomed-in version of the dashed-box region marked in the figure on the left.}}
    \label{fig:22}
\end{figure}

One of the key research topic in HEDP, especially in ICF studies, is the dynamics of instabilities. Presence of instabilities poses serious challenges to achieving high fusion gain in ICF experiments, and due to their inherent complexity, those instabilities require detailed numerical analyses. Here, SPRAY’s capability to model multidimensional instabilities is demonstrated by benchmarking 2D Rayleigh-Taylor instability simulation results against ATHENA \cite{Stone2020}. ATHENA is an astrophysical magnetohydrodynamics (MHD) code capable of both nonrelativistic and relativistic megnetohydrodynamics simulations and features adaptive mesh refinement (AMR) schemes. In the reference benchmark problem, the Atwood number is $A=\left(\rho_{u}-\rho_{l}\right)/\left(\rho_{u}+\rho_{l}\right)=1/3$, where $\rho_u$ refers to the density of the heavier fluid positioned on the upper half of the system, and $\rho_l$ corresponds to the lighter fluid on the lower half. The initial interface perturbation is defined as following:
\begin{equation}\label{3.3-70}
    y=\eta_0\cos{\left(\frac{2\pi}{\lambda}x\right)}
\end{equation}
where $\eta_0=0.01$ is the magnitude of the initial perturbation, and the wavelength $\lambda=1/3$. The simulation domain is $\left[-1/6,1/6\right]\times\left[-0.5,0.5\right]$. More information about the simulation configuration, which is identical to that of Liska et al, can be found in \cite{Liska2003}. The simulation results are shown in \hl{\mbox{\textbf{Figure \ref{fig:17}}}}, and the time evolution of the heights of the spike (heavy fluid penetrating into light fluid below) and bubble (light fluid penetrating into heavy fluid above),  along with the analytical solution from the linear theory\cite{Mikaelian2014}, are illustrated in \textbf{Figure \ref{fig:21}-\ref{fig:22}}. Here it is worthwhile to note that the detailed structures of the instability vary considerably depending on the numerical scheme. For direct comparison of simulation results among different numerical schemes, refer to Liska et al \mbox{\cite{Liska2003}}. Conventionally, the most significant physical parameters for Rayleigh-Taylor instability simulation are the growth rates of the spikes and the bubbles, and the quantitative assessment of the accuracy of SPRAY is also carried out using these parameters.

For the ATHENA run, the base resolution is $100\times300$ in $x$- and $y$-directions, respectively (\hl{see top row of} \textbf{Figure \ref{fig:17}}). For the SPRAY results, the appropriate particle spacing that corresponds to the grid size of the Eulerian grid-based ATHENA code needs to be determined.  Here, the support domain of the kernel function is matched to the grid size, i.e. $\kappa h=\Delta x_{grid}$ where $\kappa=2$ and $h$ is the smoothing length.  Due to ATHENA being a Eulerian code, the initial interface perturbation needs to be represented in a discretized manner. Depending on the resolution of the input setup, this could result in grid instabilities, which are manifested via the formation of vortices around the grid edges. In order to avoid this, density near the perturbed interface was smoothed prior to the simulation, and the same treatment was applied to the initial particle setup in SPRAY runs.

Furthermore, the differences in the qualitative aspects of the Rayleigh-Taylor instability simulations at different spatial resolutions are examined. \hl{\mbox{\textbf{Figure \ref{fig:17} (g)-(l)}}} show the simulation results with identical parameters with different spatial resolutions. The comparison of the results at various spatial resolutions attests to the SPRAY code's ability to match the capabilities of the reference code ATHENA, since the key features of ATHENA results at each resolution are also present in SPRAY's results.

The time evolution of the heights of the perturbation, which were measured as the average of the heights of the spike and the bubble, are compared in \textbf{Figure \ref{fig:21}-\ref{fig:22}}. The time axis is normalized to the growth rate, and the vertical axis is normalized to the initial perturbation magnitude. In \textbf{Figure \ref{fig:21}}, various Atwood numbers are scanned, and the results are compared with both the ATHENA results and the linear theory. For \textbf{Figure \ref{fig:22}}, the initial perturbation magnitude $\eta_0$ is scanned. From the results, it can be observed that the behavior of the average heights under various initial parameters closely follows the linear theory\cite{Taylor1950}, as well as the reference ATHENA data points. The height $\eta(t)$ of the bubble/spike in the linear phase is given as\cite{Mikaelian2014}

\begin{equation}
\eta(t)=\eta_0\cdot\cosh{(\gamma t)}=\eta_0\cdot\cosh{(\sqrt{Agk}t)}
\end{equation}

Moreover, the expected trends of the normalized heights are observed in the figures. For the Atwood number ($A$) scan, lower Atwood number (smaller density difference) leads to lower normalized heights of the spikes and the bubbles. In the case of the initial perturbation magnitude, greater $\eta_0$ causes the system to deviate from the linear regime and enter quasi- and non-linear phase eariler. This is evident from \textbf{Figure \ref{fig:22}}, since the data points for higher $\eta_0$ deviate further from the linear theory curve.

\subsection{Implosion benchmark problem - Large-scale simulation of implosion and instability dynamics}\label{sec:3.4}

Implosion dynamics is an important topic of research, especially for achieving high compression in inertial confinement fusion (ICF) research. A variety of instabilities pose difficulties in successfully compressing the deuterium-tritium (DT) fuel inside the target, and the radially converging flow dynamics enrich the physics involved in the growth of said instabilities. Here, a numerical benchmark problem originally proposed to compare Eulerian codes \mbox{\cite{Joggerst2014}} is carried out to demonstrate \hl{the} SPRAY code's capability and potential area of application.

The simulation setup is as follows: a cylindrical dense shell \hl{($\rho=1.0$ $g/cm^2$)} is filled with low density gas \hl{($\rho=0.05$ $g/cm^2$)}. The pressure of the \hl{inner gas region and the dense shell} are matched. On the outside of the cylindrical shell, a layer of high \hl{specific} internal energy \hl{($u=150$ $ergs/g$)} gas surrounds the shell. As the simulation progresses, a custom moving boundary condition is applied to the outermost section of the gas layer outside the dense shell. Both the internal energy and velocity profile of this outer section are prescribed as a function of time, which effectively drive the inward implosion. For details regarding the moving boundary condition, please refer to \mbox{\cite{Joggerst2014}}. In addition, an interface perturbation is given to the interface between the dense shell and the inner gas. The perturbation is characterized by the mode number $m$, which refers to the number of perturbation wavelengths that wrap around the cylinder:
\begin{equation}
    P(\theta) = A\cdot \cos(m\theta)
\end{equation}
where $\theta$ is the polar angle and $A$ is the perturbation magnitude.

\begin{figure}[htbp]
    \centering
    \includegraphics[width=0.8\linewidth]{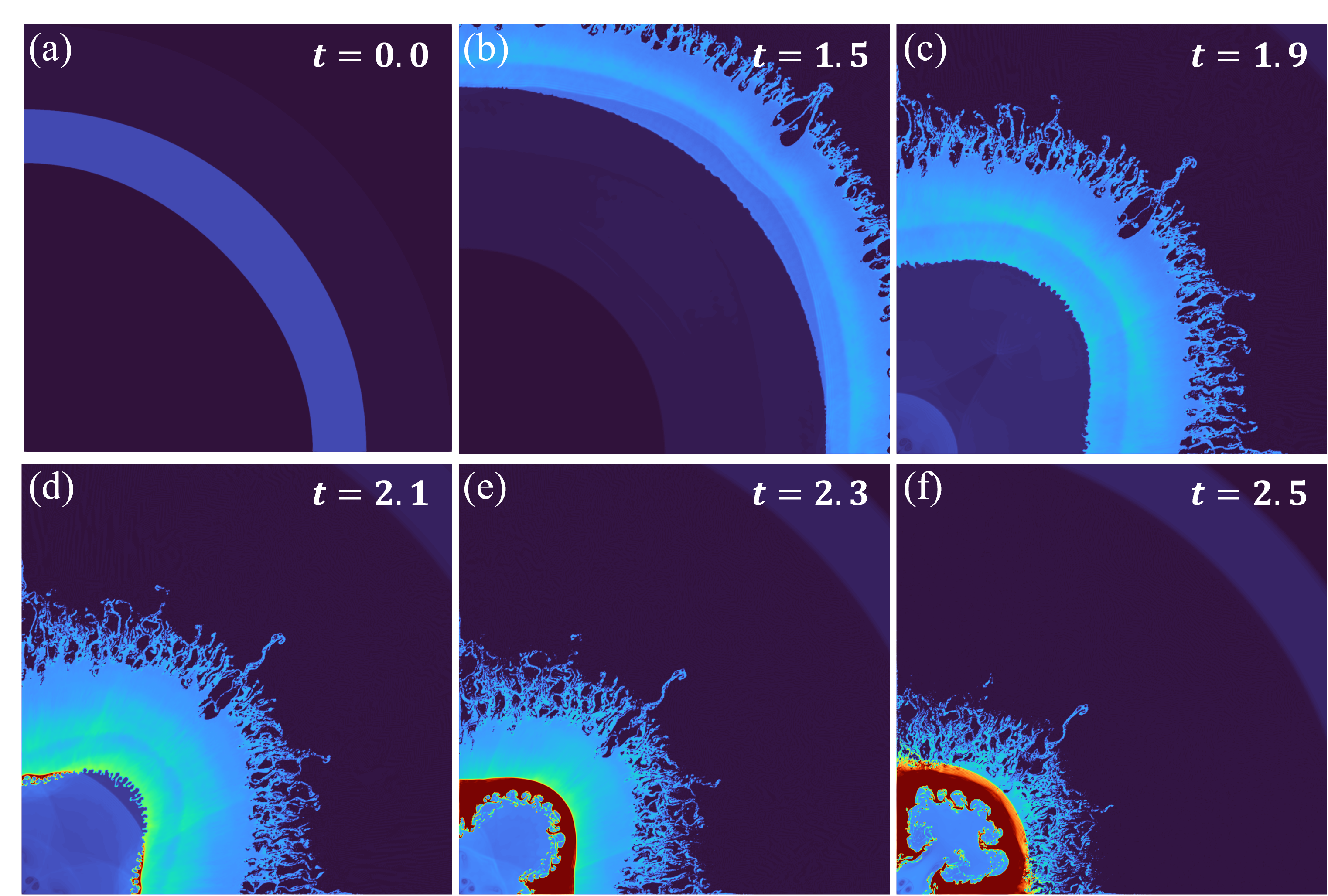}
    \caption{\textmd{$m=4$ mode simulation results. \hl{(a) shows the initial condition, and subsequent figures (b) to (f) show time progression. (b) to (f) share the same spatial scale, whereas (a) is zoomed out for illustrative purposes.} The color represents density. \hl{Simulation was ran with more than 2.8 million particles.}}}
    \label{fig:24}
\end{figure}

\begin{figure}[htbp]
    \centering
    \includegraphics[width=0.8\linewidth]{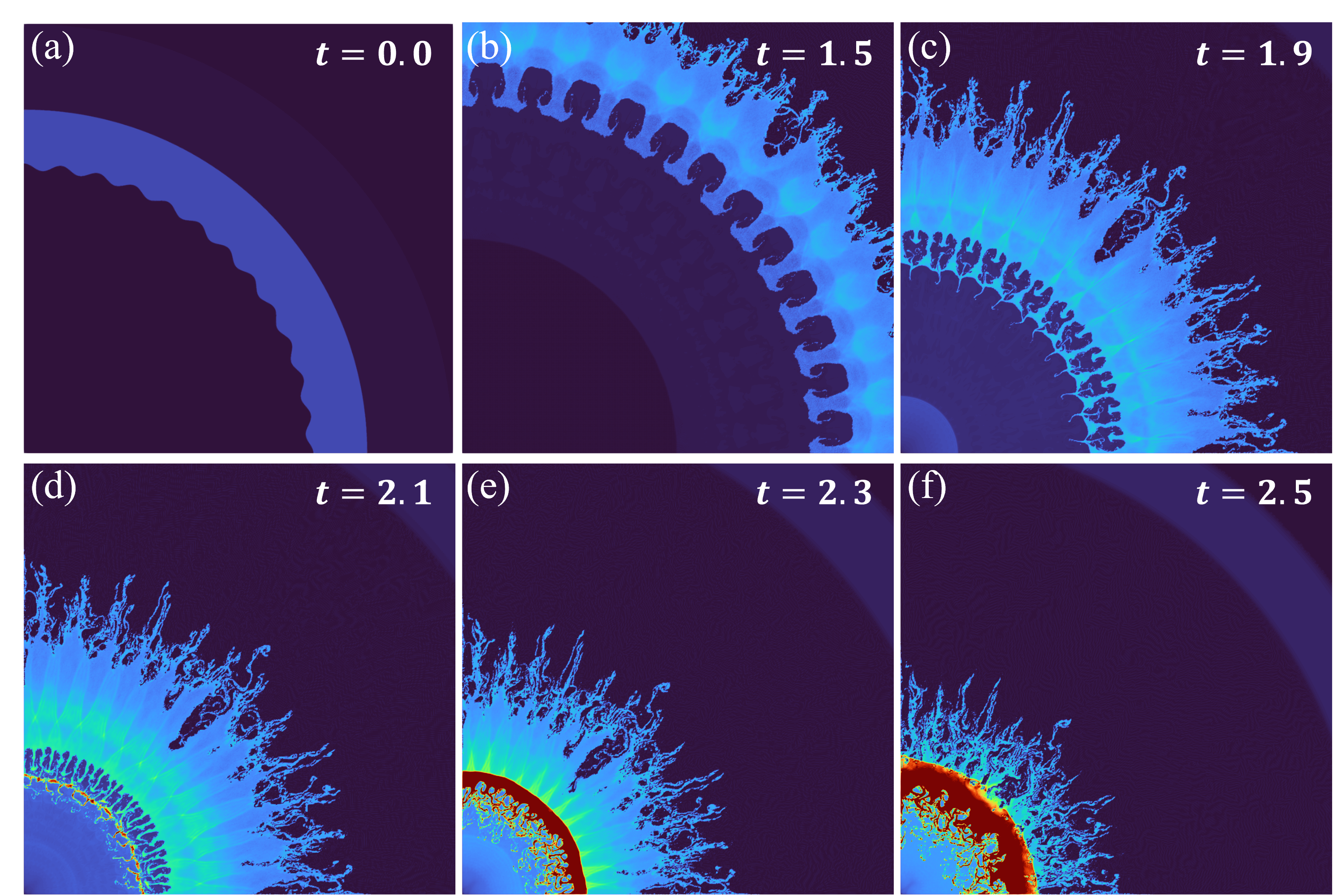}
    \caption{\textmd{$m=48$ mode simulation results. \hl{(a) shows the initial condition, and subsequent figures (b) to (f) show time progression. (b) to (f) share the same spatial scale, whereas (a) is zoomed out for illustrative purposes.} The color represents density. \hl{Simulation was ran with more than 2.8 million particles.}}}
    \label{fig:25}
\end{figure}

For the purpose of minimizing computational cost, a quadrant of the cylinder is simulated, with reflecting boundary conditions applied to the boundaries in the polar direction. Because of this setup, only mode numbers that are multiples of 4 can be properly resolved. The simulation results for \hl{the} $m=4$ and $m=48$ mode\hl{s} are presented below (see \mbox{\textbf{Figure \ref{fig:24}-\ref{fig:25}}}). In both simulations more than 2.8 million particles were used, and ideal equation-of-state is employed. \hl{These simulations took 13 hours to complete using four NVIDIA A100 GPUs.}

The simulation results shown in \hl{\mbox{\textbf{Figure \ref{fig:24}}}} exhibit all of the expected physical phenomena. The formation of spikes and bubbles in accordance with the mode number is clear in \hl{\mbox{\textbf{Figure \ref{fig:24}(f)}}}. For the high mode number case ($m=48$), the spike and bubble formation occurs in \hl{an} earlier time slice (\hl{\mbox{\textbf{Figure \ref{fig:25}(b)}}}), which is in agreement with the reference literature \mbox{\cite{Joggerst2014}}. In addition, the turbulent mixing in the late phase of high mode number instability is well resolved in \hl{\mbox{\textbf{Figure \ref{fig:25}(f)}}}. These results demonstrate the robustness and capability of \hl{the} SPRAY code in simulating complex implosion and instability dynamics. The detailed progression of the simulation is shown in \mbox{\textbf{Figure \ref{fig:24}-\ref{fig:25}}}.

\subsection{Laser driven inertial confinement fusion target compression simulation - Mesh-free laser ray-tracing in arbitrary geometry}\label{sec:3.5}

One of the most popular topic\hl{s} in \hl{the} high-intensity laser-plasma interaction research is the laser-driven inertial confinement fusion (ICF). By irradiating \hl{the} ablator material with high power laser\hl{s}, the ablation pushes the DT fuel inward, causing the target capsule to implode. Therefore, the coupling between \hl{the} laser energy deposition routine \hl{and} the plasma simulation is critical to \hl{the} successful numerical \hl{modeling} of this \hl{experiment}. Here, a laser driven compression of a ICF target is simulated with SPRAY to verify its laser ray-tracing scheme and its integration to the code.

The simulation setup is similar to that of ref. \mbox{\cite{Atzeni2004}}. A thin cylindrical shell of DT ice with inner radius of 1.76 mm and outer radius of 1.934 mm is coated with a ablator layer composed of polystyrene (CH) with a thickness of 0.037 mm. The interior of the shell is assumed to be filled with low density DT gas. \hl{A} 250 nm wavelength \hl{laser} is set to uniformly irradiate the target at 500 TW of power, and 2048 rays are used in total. Because the nuclear fusion reaction of DT is not modelled in this code, the simulation is set to terminate before the deceleration phase begins. \hl{The option to solve radiation transport was disabled in this simulation. This is because, although radiation transport plays an important role in the latter stages of the inertial confinement fusion process, its effects are not as pronounced in the earlier phases. For this reason, computational efficiency was prioritized over strict physical accuracy.} Tabulated equation-of-state data of both DT and CH were generated from MPQEOS code. The implicit heat transport module with BiCGSTAB solver described in \mbox{\textbf{Section \ref{sec:2.1.2}}} is used.

\begin{figure}[htbp]
    \centering
    \includegraphics[width=1.0\linewidth]{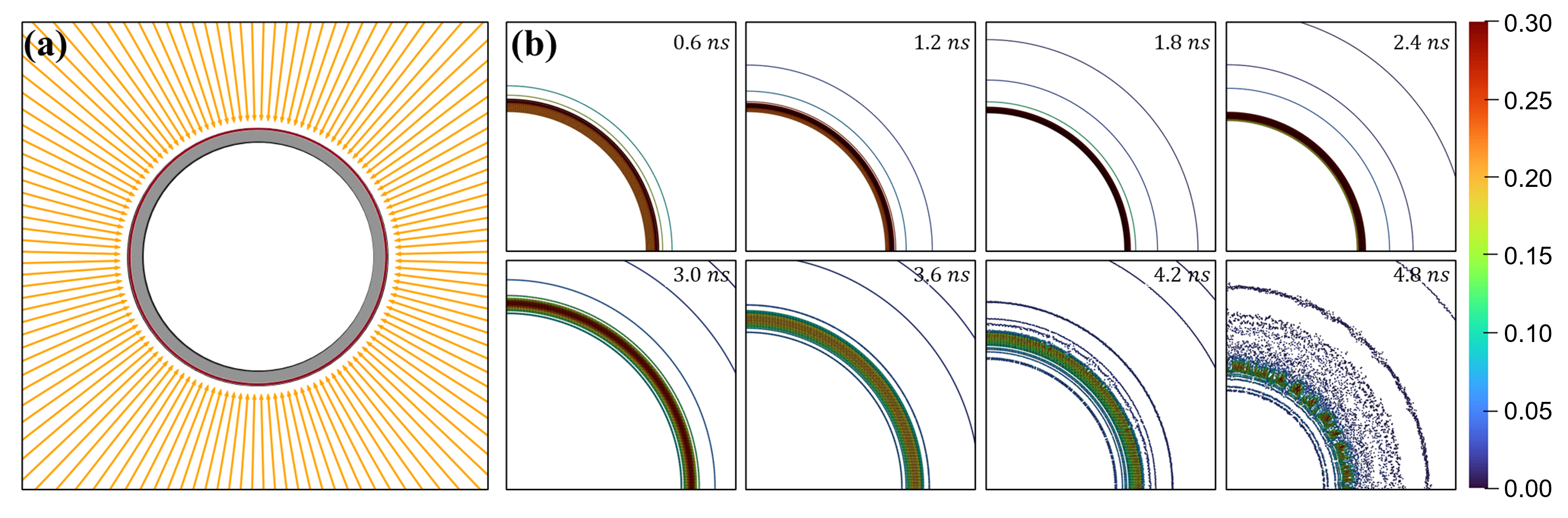}
    \caption{\textmd{Laser-driven implosion simulation snapshots. The target is composed of polystrene (CH) ablator shell (red layer) and DT ice inner layer (gray layer). 2,048 laser rays are uniformly incident on the target, with the total power of 500 TW and wavelength of 250 nm. (a) shows the initial setup of the simulation, and the orange lines indicate laser rays. The number of rays drawn is reduced by a factor of 16 for better visibility. (b) shows the time evolution of the simulation. The upper four figures show the ablation of the outer polystyrene layer, and the lower four figures (zoomed in), show the implosion of the DT shell. The color corresponds to the density of each particle \hl{in the unit of $g/cm^3$}.}}
    \label{fig:26}
\end{figure}

\begin{figure}[htbp]
    \centering
    \includegraphics[width=0.6\linewidth]{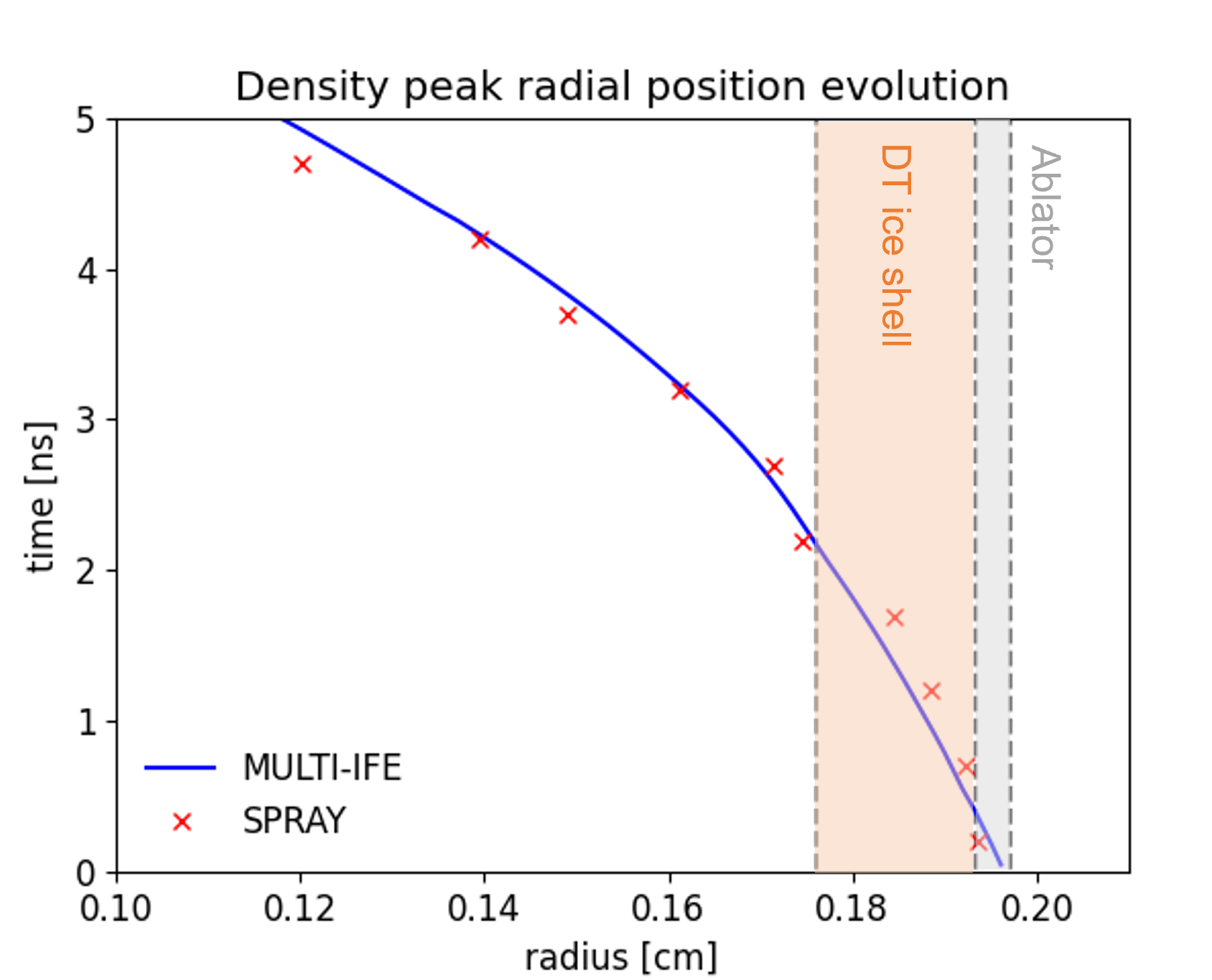}
    \caption{\textmd{The radial position of density peak from SPRAY simulation is compared with a reference code MULTI-IFE\cite{Ramis2016}. The blue line is the reference data, and the red crosses are measured from SPRAY results. The vertical dotted lines indicate the initial position of the ablator and the DT ice shell.}}
    \label{fig:27}
\end{figure}

The results of the simulation are shown in \mbox{\textbf{Figure \ref{fig:26}}}. As the simulation progresses, the inward implosion of the DT shell due to ablation of the outer CH layer can be observed. It is worth noting that, thanks to the random sampling of ray launch position (described in \mbox{\textbf{Section \ref{sec:2.5}}}), there is no immediate formation of instabilities due to non-uniform laser irradiation. In addition, the free surface boundary treatment scheme described in \mbox{\textbf{Section \ref{sec:2.2}}} is applied, so the rapid ablation does not distort the density or pressure profiles. The physical validity of the result is verified using a reference code MULTI-IFE \mbox{\cite{Ramis2016}}, which yields the implosion velocity for a perfectly uniform implosion case via one-dimensional simulation. When the radial movement of the density peak is tracked and plotted over time (see \mbox{\textbf{Figure \ref{fig:27}}}), the results of the two codes are in good agreement, which indicates that the coupling of laser energy to the plasma is accurate and physically sound.

\section{Conclusions}
In this paper, a SPH-based RHD code SPRAY is presented. It is a particle-based, mesh-free, Lagrangian code intended to simulate long pulse high-intensity laser-plasma interaction in HEDP regime. It utilizes GPU for its massively parallel approach, and its SPH foundation gives it an edge on traditionally challenging issues such as tracking fluid surfaces and simulating complex deformations, which usually necessitated heavily time-consuming numerical techniques. \hl{An algorithm suitable for GPU parallelization acceleration is implemented in the nearest neighbor particle search (NNPS) routine for optimal performance.} The SPH formulations are tailored toward simulating laser-target interaction, thus yielding accurate and reliable results. \hl{A novel particle-based laser ray-tracing scheme is also developed with integration with the SPH foundation in mind.} The hydrodynamics governing equations are solved in explicit scheme, whereas the radiation transport is handled in implicit manner.

The versatility and reliability of the code is verified with a series of benchmarks. First, the validity of the hydrodynamics calculation as well as the ability to capture and resolve shock waves is confirmed with the Sod shock tube benchmark as the simulation results are in good agreement with the analytical solution. Second, the radiation transport routine and the coupling of laser energy to the plasma are validated by comparing the simulation results with a well-established code MULTI-IFE\mbox{\cite{Ramis2016}}. The density, velocity, ion/electron temperature profile evolution calculated by SPRAY closely matched that of the reference code. Third, the code’s ability to model multidimensional instabilities was examined with the classical Rayleigh-Taylor instability benchmark. The growth rate of the spike and the bubble was compared with that of a reference code ATHENA\mbox{\cite{Stone2020}}, along with the linear theory predictions. The results from SPRAY are shown to exhibit the initial linear phase growth, and are successful in reproducing the reference code’s results at various resolutions. Fourth, simulation of implosion of an ICF target is carried out to illustrate a potential application of SPRAY code. The expected growth of the instabilities as well as turbulent mixing phenomena are clearly observed, thus demonstrating the robustness of the code. Fifth, a laser-driven implosion simulation results are shown, which exhibiting the capability of full multidimensional laser ray-tracing routine coupled with the code. Only two-dimensional results are included in this study, \hl{as the extension to three dimensions is left as a future work.}

\hl{There are several clear limitations of the SPRAY code. First, there are inherent restrictions stemming from using SPH methods in the context of high energy density physics. Drastic changes in the state of matter, i.e. phase change from solids to plasmas via laser ablation, are notoriously challenging to model. Thanks to years of effort and research, Eulerian approaches have more or less matured in terms of modeling such phase changes, but it remains one of the grand challenges of SPH.

Second, there is room for improvement of the implementation of radiation transport solver. As stated in this paper, SPRAY code solves the radiation transport equations derived with the gray approximation. The physical accuracy of the model would certainly improve if the code were to be extended to support multigroup approximation, especially when modeling indirectly driven inertial confinement fusion where strong X-rays act as the driver. Moreover, although the flux-limited diffusion approach yields reasonable results in both the optically thick and thin limits, it breaks down in the intermediate regime. A more realistic modeling could be achieved with improved methods such as the variable Eddington method\mbox{\cite{Petkova2010}}. Furthermore, a more modern numerical solver could increase both the performance and scalability of the Jacobi-based implicit solver implemented in the code. It is expected that a more computationally efficient scheme could dramatically reduce the number of iterations required for convergence by orders of magnitude.
}

To the authors’ knowledge, this is the first attempt to apply SPH techniques to HEDP research involving laser-target interactions. Therefore, the significance of this research lies in the fact that it opens possibility of tackling challenging issues in the HEDP field with the rapidly advancing numerical analysis capabilities and techniques of SPH. Works in progress that would be implemented in future releases include laser beam-beam interaction \hl{modeling}, as well as high-performance and highly optimized multi-group diffusion approximation radiation transport module. These features, if implemented properly, would enable SPRAY to address its current limitations and be applicable to even broader range of HEDP numerical studies.

\section*{CRediT authorship contribution statement}
\textbf{Min Ki Jung:} Conceptualization, Methodology, Software, Validation, Investigation, Writing - Original Draft, Writing - Review \& Editing, Visualization \textbf{Hakhyeon Kim:} Resources, Investigation \textbf{Su-San Park:} Methodology, Software \textbf{Eung Soo Kim:} Conceptualization, Methodology, Software \textbf{Yong-Su Na:} Conceptualization, Writing - Review \& Editing \textbf{Sang June Hahn:} Conceptualization, Methodology, Validation, Resources, Writing - Review \& Editing, Supervision, Project administration

\section*{Declaration of competing interest}
The authors declare that they have no known competing financial interests or personal relationships that could have appeared to influence the work reported in this paper.

\section*{Acknowledgments}
This work was supported by the Defense Research Laboratory Program of the Defense Acquisition Program Administration and the Agency for Defense Development of the Republic of Korea. We gratefully acknowledge The Research Institude of Energy and Resources and The Institute of Engineering Research at Seoul National University.

%\section*{References}

%%Vancouver style references.
\bibliographystyle{model1-num-names}
\bibliography{refs}

\end{document}